\documentstyle[aps,pre,twocolumn,epsf,floats]{revtex}
\begin{document}
\def\OP {\tensor P}
\def\B.#1{{\bf {#1}}}
\renewcommand{\thesection}{\arabic{section}}
\title{{\rm PHYSICAL REVIEW E   \hfill  Version of \today}\\~~\\
  Computing the Scaling Exponents in Fluid Turbulence from First
  Principles:  \\ the Formal Setup } 
\author {Victor S. L'vov and Itamar  Procaccia} 
\address{Department of~~Chemical Physics, The Weizmann
  Institute of Science, Rehovot 76100, Israel} 
\maketitle
\begin{abstract}
  We propose a scheme for the calculation from the Navier-Stokes
  equations of the scaling exponents $\zeta_n$ of the $n$th order
  correlation functions in fully developed hydrodynamic turbulence.
  The scheme is nonperturbative and constructed to respect the
  fundamental rescaling symmetry of the Euler equation. It constitutes
  an infinite hierarchy of coupled equations that are obeyed
  identically with respect to scaling for any set of scaling exponents
  $\zeta_n$.  As a consequence the scaling exponents are determined by
  solvability conditions and not from power counting.  It is argued
  that in order to achieve such a formulation one must recognize that
  the many-point space-time correlation functions are not scale
  invariant in their time arguments.  The assumption of full scale
  invariance leads unavoidably to Kolmogorov exponents.  It is argued
  that the determination of of all the scaling exponents $\zeta_n$
  requires equations for infinitely many renormalized objects. One can
  however proceed in controlled successive approximations by
  successive truncations of the infinite hierarchy of equations.Clues
  as to how to truncate without reintroducing power counting can be
  obtained from renormalized perturbation theory. To this aim we show
  that the fully resummed perturbation theory is equivalent in its
  contents to the exact hierarchy of equations obeyed by the $n$th
  order correlation functions and Green's function. In light of this
  important result we can safely use finite resummations to construct
  successive closures of the infinite hierarchy of equations.  This
  paper presents the conceptual and technical details of the scheme.
  The analysis of the high-order closure procedures which do not
  destroy the rescaling symmetry and the actual calculations fWork4or
  truncated models will be presented in a forthcoming paper in
  collaboration with V. Belinicher.
\end{abstract}
\section{Introduction}
The aim of this paper is to present a general scheme for the
calculation of the scaling exponents characterizing the statistical
quantities that arise in the description of fully developed
hydrodynamic turbulence.  These statistical quantities are various
averages computed from the fundamental field in hydrodynamics,
the velocity field of the fluid. Denote the Eulerian velocity field
as ${\B.u}({\B.r},t)$ where $\B.r$ is a point in $d$-dimensional space
(usually $d=2$ or 3) and $t$ is the time. Statistical quantities that
have attracted years of experimental and theoretical attention
\cite{MY,96Sre,94Nel,Fri} are the structure functions of velocity
differences, denoted as $S_n(R)$
 \begin{equation}
 S_n(R) = \left\langle\vert\hbox{\B.u}({\B.
 r}+{\B.R},t) -{\B.u}({\B.r},t)\vert^n\right\rangle \ ,
 \label{a5}
 \end{equation} 
 where $\left<\dots\right>$ stands for a suitably defined ensemble
 average.  It has been asserted for a long time that the structure
 functions scale as a function of the separation $R$ according to
\begin{equation}
 S_n(R) \sim R^{\zeta_n}  \ ,        
    \label{scalelaw}                
 \end{equation}
 where $\zeta_n$ are known as the scaling exponents of the structure
 functions.  It is assumed here that the separation $R$ lies in the
 so-called ``inertial range", i.e.  $\eta \ll R \ll L$ with $\eta$
the inner viscous scale and $L$ the outer integral scale
 of turbulence. One of the major questions in fundamental turbulence
 research is whether these scaling exponents are correctly predicted
 by the classical Kolmogorov 41 theory in which $\zeta_n = n/3$, or
if these exponents manifest the phenomenon of ``multiscaling"
 with $\zeta_n$ a nonlinear function of $n$, as has been
 indicated by physical and numerical experiments
 \cite{cil,tab,sre,chen,pras}.
 
In an attempt to develop a consistent theory of turbulence one may
define statistical quantities that depend on many spatial and
 temporal coordinates. Defining the velocity difference
 ${\B.w}({\B.r},{\B.r}',t)$ as
\begin{equation}
{\B.w}({\B.r},{\B.r}',t) \equiv {\B.u} ({\B.r}',t)-
 {\B.u}({\B.r},t) \ , \label{diff}
\end{equation}
one considers the $n$-rank tensor space-time correlation function
\begin{eqnarray}
&&\B.F_n({\B.r}_1,{\B.r}'_1,t_1;{\B.r}_2,{\B.r}'_2,t_2;
\dots;\B.r_n,{\B.r}'_n,t_n)
\nonumber \\
&=& \left< \B.w({\B.r}_1,{\B.r}'_1,t_1) 
\B.w({\B.r}_2,{\B.r}'_2,t_2) \dots
\B.w({\B.r}_n,{\B.r}'_n,t_n) \right> \ . \label{defFt}
\end{eqnarray}
The simultaneous correlation function $\B.T_n$ is obtained from
${\B.F}_n$ when $t_1=t_2\dots=t_n$. In statistically stationary
turbulence the equal time correlation function is time independent,
and we denote it as
\begin{eqnarray}
&&{\B.T}_n({\B.r}_1,{\B.r}'_1;{\B.r}_2,{\B.r}'_2;
\dots;{\B.r}_n,{\B.r}'_n)
\nonumber \\
&=& \left< \B.w({\B.r}_1,{\B.r}'_1,t)
\B.w({\B.r}_2,{\B.r}'_2,t) \dots
\B.w({\B.r}_n,{\B.r}'_n,t) \right> \ .
\label{defF}
\end{eqnarray}
One expects that when all the separations $R_i\equiv|\B.r_i-\B.r'_i|$
are in the inertial range, $\eta\ll R_i\ll L$, the simultaneous
correlation function is scale invariant in the sense that
\begin{eqnarray}
&&{\B.T}_n(\lambda {\B.r}_1,\lambda{\B.r}'_1;\lambda{\B.r}_2,
\lambda{\B.r}'_2;
\dots;\lambda{\B.r}_n,\lambda{\B.r}'_n)
\nonumber \\
&=& \lambda^{\zeta_n}{\B.T}_n({\B.r}_1,{\B.r}'_1;{\B.r}_2,
{\B.r}'_2;
\dots;{\B.r}_n,{\B.r}'_n) \ , \label{scaleinv}
\end{eqnarray}
and the exponent $\zeta_n$ is numerically the same as the one
appearing in Eq.~(\ref{scalelaw}).

One of the major difference between the study of statistical
turbulence and other examples of anomalous scaling in physics (like
critical phenomena) is that there is no theory for the simultaneous
correlation functions (\ref{defF}) that does not involve the many time
correlation functions (\ref{defFt}). {\em Turbulence is a truly
  dynamical problem}, and there is no free energy functional or a
Boltzmann factor to provide a time-independent theory of the
statistical weights. The theory for the simultaneous quantities
(\ref{defF}) involves {\em integrals over the time variables of the many time quantities} (\ref{defFt}). One must therefore learn how to
perform these time integrations properly.

We propose here a point of departure from all previous attempts to
compute the anomalous exponents $\zeta_n$ from first principles, based
on the understanding that these functions are not scale invariant in
their time arguments; this allows us to build a new approach.  Naively
one could assume that the scale invariance property extends to the
time correlation functions in the sense that
\begin{eqnarray}
&&{\B.F}_n(\lambda {\B.r}_1,\lambda{\B.r}'_1,\lambda^{z_n}t_1;
\dots;\lambda{\B.r}_n,\lambda{\B.r}'_n,\lambda^{z_n}t_n)
\nonumber \\
&=& \lambda^{\zeta_n}{\B.F}_n({\B.r}_1,{\B.r}'_1,t_1;
\dots;{\B.r}_n,{\B.r}'_n,t_n) \ , \quad({\rm not~true!})\ , 
\label{tscaleinv}
\end{eqnarray}
with a dynamical scaling exponent $z_n$ which can be $n$ dependent.
We have shown \cite{97LPP} that this is not the case, and in this
paper we explain that using an adequate representation of the time
dependence {\em leads to a calculation scheme for the scaling
exponents}. We will reiterate that even in fully resummed theories,
if we assume scale invariance in time we lose the availability of
anomalous scaling in favor of K41 scaling. Full respect to the
dynamical nature of the problem is required in order to proceed in
understanding anomalous scaling.

Our strategy in developing a scheme of computation of the scaling
exponents can be described as follows: The first step is to transform
the Navier-Stokes equations using the Belinicher-L'vov transformation
\cite{87BL}, see Eqs.(\ref{a2})-(\ref{newBL}). The resulting velocity
field is denoted $\B.{\cal W}$, and the correlation function over
these velocity fields is denoted $\B.{\cal F}_n$, see
Eq.~(\ref{defFtime}).  Next we construct the infinite hierarchy of
equations that relate the rate of change of the $n$th order
correlation function $\B.{\cal F}_n$ to a space integral over a vertex
convolved with the ($n+1$)th order correlation function $\B.{\cal
  F}_{n+1}$. This hierarchy is given by
Eqs.(\ref{balt}),(\ref{baltver}) below and it is exact. Next we
explain, using the fusion rules \cite{LPfuse,97LP} that govern the
asymptotic properties of the correlations functions when groups of
their spatial arguments coalesce together, that the spatial integrals
converge both in the UV and the IR limits. This result allows us to
show that if Eq.~(\ref{tscaleinv}) were true, the hierarchy of
equations could be studied by power counting, and the only solution
would be linear (i.e. $\zeta_n$ linear in $n$) rather than anomalous
scaling. We will show that Eq.~(\ref{tscaleinv}) is not true, and that
the many-time correlation functions must be ``multiscaling" in their
time representation. This fact means that the hierarchy of equations
cannot be studied by power counting. The next step is to represent the
space and time dependent correlation functions in a way that exposes
their multiscaling characteristics \cite{97LPP}.  This form
(\ref{repres}) is a convenient form that makes use of the
``multifractal" representation, which was found useful before in the
context of the simultaneous objects \cite{Fri}.  This representation
amounts to saddle point integrals over correlation functions that
respect the exact symmetry of the Euler equation, i.e.  the symmetry
of rescaling according to
\begin{equation}
\B.r \to \lambda \B.r, \quad \B.w \to \lambda^h \B.w,
 \quad t\to \lambda^{1-h}  t \ . \label{ressym}
\end{equation}
Using this representation we will see that the hierarchy of equations
for the correlation functions is satisfied in the sense of power
counting for any value of $h$, cf. Eq.~(\ref{greater}).  The
information about the scaling exponents is hidden in the coefficients
of the equation but cannot be read from power counting.

We are still faced with an infinite hierarchy of equations. This is
not accidental; the calculation of infinitely many anomalous exponents
in turbulence does indeed necessitate a calculation of infinitely many
renormalized objects. Nevertheless we want to achieve a finite,
approximate calculation of lower order exponents (i.e $\zeta_n$ with
for the first integer values $n$).  To this aim we need to truncate
the infinite hierarchy of equations. The way to achieve an intelligent
truncation can be seen from the study of renormalized perturbation
theory \cite{P-1}. Perturbative theories of turbulence have fallen
into some ill-repute since the problem of turbulence does not have any
small parameter that guarantees convergence of perturbative
expansions. To increase our confidence in this approach, we first
prove that the content of fully resummed perturbation theory is
identical to the exact hierarchy of equations for $n$-order
correlation functions and Green's functions. This important result,
which is demonstrated here for the first time, allows us to proceed
safely in using partially resummed perturbation theory to guide our
truncation of the hierarchy of exact equations. The clue is that we
want to preserve the invariance of the theory under the rescaling
(\ref{ressym}). We outline how to do this in Section 8 and Appendix A
of this paper. The actual analysis of the method of truncating while
preserving the symmetry, and an approximate calculation of the low
order exponents will be presented in a forthcoming paper \cite{97BLP}.
Here we present the conceptual steps and technical issues.
\section
{Nonperturbative Formulation of the Statistical Theory of Turbulence}
In order to derive anomalous exponents it is essential to develop a
theory that is nonperturbative. In addition, one needs to deal with
the effect of the sweeping of small scales by large scale motions.
This effect is only kinematic, but it can mask the inherent time
scales associated with nonlinear interactions, and therefore has to be
carefully taken into account. To this aim we start this section with a
short review of the equations of motion in the Belinicher-L'vov
representation (subsection A), followed by the introduction of the
statistical quantities of interest, which are $n$-order space time
correlation functions and Green's functions in subsection B. In
subsection C we derive the exact hierarchy of equations satisfied by
these quantities, the main results being Eqs.~(\ref{balt})
(\ref{baltver}) and (\ref{balGn}).
\subsection{Equations of motion} 
The analytic theory of turbulence is based on the Navier-Stokes
equation for the Eulerian velocity field ${\B.u}({\B.r},t)$. In the
case of an incompressible fluid they read
 \begin{equation}
  \partial   {\B.u}/ \partial t + (   {\B.u}\cdot 
  {\bbox{\nabla}})   {\B.u} -\nu\nabla^2   {\B.u} 
 +{\bbox{\nabla }} p = {\B.f}, \quad {\bbox{\nabla}}\cdot 
  {\B.u}=0\,,
 \label{b1}
 \end{equation}
 where $\nu$ is the kinematic viscosity, $p$ is the pressure, and
 ${\B.f}$ is some forcing which maintains the flow. Since we are
 interested in incompressible flows, we project the longitudinal
 components out of the equations of motion. This is done with the help
 of the projection operator $\tensor P$ which is formally written as
 $\tensor P$ $\equiv -\nabla^{-2}{\bbox{\nabla }}\times {\bbox{\nabla
     }}\times$. Applying $\OP$ to Eq.~(\ref{b1}) we find
 \begin{equation}
 (\partial/ \partial t-\nu\nabla^2){\B.u}+
 \OP
 ({\B.u}\cdot\bbox{\nabla}){\B.u}
 =\,\OP
 {\B.f}\ .
 \label{b6}
 \end{equation}
 This equation has been used as a starting point for a field-theoretic
 perturbation theory, in which the nonlinear term acts as a
 ``perturbation" on the linear part of the equation. Such an approach
 is fraught with difficulties simply because the natural statistical
 objects which appear in the perturbation
 expansions\cite{61Wyl,73MSR,76Dom,76Jan} of the Navier-Stokes
 equation are correlation functions $ \tilde F^{\alpha\beta}({\bf
   r},{\B.r}',t,t')$ of the velocity field ${\B.u}({\B.r},t)$ itself:
 \begin{equation} 
\tilde F^{\alpha\beta}({\B.r},{\B.r}',t,t') 
= \left\langle u^\alpha({\B.r},t)
 u^\beta({\B.r}',t')\right\rangle \ .
 \label{a1}
 \end{equation}
 The problem is that the correlator $\tilde
 F^{\alpha\beta}({\B.r},{\B.r}',t,t')$ is not universal, since it is
 dominated by contributions to $u_\alpha({\B.r},t)$ which come from
 the largest scales in the fluid flow, and these are determined by the
 features of the energy injection mechanisms. This physical fact is
 reflected in the theory as infra-red divergences that have plagued
 the development of analytic approach for decades. Indeed, all the
 early attempts to develop a consistent analytic approach to
 turbulence, notably the well known theories of Wyld\cite{61Wyl} and
 of Martin, Siggia and Rose\cite{73MSR}, shared this problem. One
 needs to transform Eq.~(\ref{b1}) such that the statistical objects
 that appear naturally will be written in terms of velocity
 differences which can be universal. We like to use the
 Belinicher-L'vov (BL) transformation for this purpose, but any other
 transformation that reaches the same goal is equally acceptable.  We
 will stress that the results that we obtain below do not depend on
 the details of the transformation, and we employ the BL
 transformation because it is particularly straightforward.  The
 equation of motion that is obtained from the BL-transformation is
 exact, and there is no approximation involved in this step. In terms
 of the Eulerian velocity ${\B.u}({\B.r},t)$ Belinicher and L'vov
 defined the field ${\B.v}({\B.r}_0,t_0\vert {\B.r},t)$ as
 \begin{equation}
 {\B.v}({\B.r}_0,t_0\vert {\B.r},t)\equiv {\B.u}\lbrack{\B.r}
 +\mbox{\boldmath$\rho$}
 ({\B.r}_0,t),t\rbrack
 \label{a2}
 \end{equation}
 where
 \begin{equation}
 \mbox{\boldmath$\rho$}  ({\B.r}_0,t)
  =\int_{t_0}^{t} ds {\B.u }[{\B.r}_0 +\mbox{\boldmath$\rho$}({\B.
  r}_0,s) ,s] \ .  
  \label{a3} 
 \end{equation}
 Note that $\mbox{\boldmath$\rho$} ({\B.r}_0,t)$ is precisely the
 Lagrangian trajectory of a fluid particle that is positioned at ${\bf
   r}_0$ at time $t=t_0$. The field ${\B.v}({\bf r}_0,t_0\vert
 {\B.r},t)$ is simply the Eulerian field in the frame of reference of
 a single chosen material point $\B.\rho(\B.r_0,t)$.  The observation
 of Belinicher and L'vov \cite{87BL} was that the variables $\B.{\cal
   W}(\B.r_0,t_0|\B.r,\B.r',t)$ defined as
 \begin{equation}
\B.{\cal W}(\B.r_0,t_0|\B.r,\B.r',t)\equiv   
\B.v( {\B.r}_0,t_0| {\B.r},t)-
\B.v( {\B.r}_0,t_0| {\B.r}',t)  \ , \label{newBL}
\end{equation}
exactly satisfy a Navier-Stokes-like equation in the incompressible
limit.  Consequently one can develop a diagrammatic perturbation
theory in terms of these variables \cite{P-1}. The resulting theory is
free of the two related problems that we discussed above: the
$\left\langle\B.{\cal W}\B.{\cal W}\right \rangle$ correlators are
universal for $\vert{\B.r}-{\B.r}'\vert$ in the inertial range, and
the theory is free of infra-red (and ultra-violet) divergences
resulting from sweeping. It can be also shown \cite{87BL,P-1} that
Kolmogorov 1941 scaling is an order by order solution of the resulting
theory.  This formulation has two important properties. (i) The
simultaneous correlators of
${\B.v}\lbrack\hbox{\B.r}_0,t_0\vert\hbox{\B.r}, t\rbrack$ are
$(\B.r_0,t_0)$-independent and identical to the simultaneous
correlators of $\hbox{\B.u}(\hbox{\B.r},t)$. The reason is that for
stationary statistics the simultaneous correlators of an arbitrary
number of factors of ${\B.v}\lbrack\hbox{\bf
  r}_0,t_0\vert\hbox{\B.r},t\rbrack$ does not depend on $t$, and in
particular one can take $t=t_0$. The property then follows directly
from Eqs.~(\ref{a2})-(\ref{a3}). (ii) The correlators of $\B.{\cal W}$
are closely related to the structure functions of {\B.u},
Eq.~(\ref{a5}).  Clearly
 \begin{equation} S_n(\hbox{\B. r}-\hbox{\B.r}') 
= \left\langle\vert\B.{\cal W}(\B.r_0,t_0|\B.r,\B.r',t)
\vert^n\right\rangle \ .
 \label{a6}
 \end{equation} 
 As the structure functions of the Eulerian velocity differences have
 for years been at the focus of experimental research, the formulation
 in terms of the variables $\B.{\cal W}$ gives a direct link between
 theory and experiments.  We will refer to the variable $\B.{\cal
   W}(\B.r_0,t_0|\B.r,\B.r',t)$ as the BL velocity differences.  We
 now perform the Belinicher-L'vov change of variables
 (\ref{a2}-\ref{a3}) together with ${\B.f}[{\B.r}
 +\mbox{\boldmath$\rho$}({\B.r}_0,t_0,t),t] ={\mbox{\boldmath$\phi$}}
 \lbrack{\B.r}_0,t_0\vert {\bf r},t\rbrack$.  Using the Navier Stokes
 equation and the chain rule of differentiation we find the equation
 of motion for $\B.{\cal W}(\B.r_0,t_0|\B.r,\B.r',t)$: \FL
\begin{eqnarray}
&&\Big[{\partial\over \partial t}+\hat{\cal L}
-\nu (\nabla_r^2+\nabla_r'^2)\Big]
\B.{\cal W}(\B.r_0,t_0|\B.r,\B.r',t)\nonumber \\
&&=\tensor{\B.P}\B.\phi(\B.r_0,t_0|\B.r,t)-
\tensor{\B.P}'\phi(\B.r_0,t_0|\B.r',t). \label{newNS}
\end{eqnarray}
We introduced an operator $ \hat{\cal L}=\hat{\cal
  L}(\B.r_0,t_0|\B.r,\B.r',t)$ as follows:
\begin{eqnarray}
 \hat{\cal L}(\B.r_0,t_0|\B.r,\B.r',t)&\equiv&
\tensor{\B.P} \B.{\cal W}({\B.r}_0,t_0|{\B.r},\B.r_0,t)
\cdot{\B.\nabla_r}
\nonumber\\ &+&
\tensor{\B.P}' \B.{\cal W}({\B.r}_0,t_0|{\B.r'},\B.r_0,t)
\cdot{\B.\nabla_r'}\ .
\label{calL}
 \end{eqnarray}
 We remind the reader that the application of ${\tensor{\B.P} }$ to
 any given vector field $ {\B.a}( {\B.r})$ is non local, and has
 the form:
 \begin{equation}
 \lbrack\tensor{\B.P}{\B.a}(
 {\B.r})\rbrack^\alpha =\int d  \tilde\B.r P^{\alpha\beta}(
 {\B.r}-\tilde\B.r)a^\beta(\tilde\B.r).
 \label{b2}
 \end{equation}
 The explicit form of the kernel can be found, for example, in
 \cite{MY}.  In (\ref{calL}) $\tensor{\B.P}$ and $\tensor{\B.P}'$ are
 projection operators which act on fields that depend on the
 corresponding coordinates $\B.r$ and $\B.r'$. The equation of motion
 (\ref{newNS}) forms the basis of the following discussion of the
 statistical quantities. It has the tremendous advantage over the
 Eulerian version (\ref{b6}) in that all the sweeping effects are
 removed explicitly.  We note that the equation of motion is
 independent of $t_0$, and from now on we drop the argument $t_0$ in
 $\B.{\cal W}(\B.r_0,t_0|\B.r,\B.r',t)$.
\subsection{The statistical quantities}
The fundamental statistical quantities in our study are the
many-time, many-point, ``fully-unfused", $n$-rank-tensor
correlation function of the BL velocity differences $\B.{\cal
  W}_j\equiv \B.{\cal W} (\B.r _0|\B.r_j \B.r'_j, t_j)$. To simplify
the notation we choose the following short hand notation:
\begin{equation}
X_j\equiv \{\B.r_j,\B.r'_j,t_j\}, \  x_j
\equiv \{\B.r_j,t_j\}, \  \B.{\cal W}_j\equiv
\B.{\cal W}(X_j) \ , \label{short}
\end{equation}
\begin{equation}
\B.{\cal F}_n(\B.r_0|X_1 \dots X_n)
= \left< \B.{\cal W}_1 \dots \B.{\cal W}_n \right> \ .
\label{defFtime}
\end{equation}
By the term ``fully unfused" we mean that all the coordinates are
distinct and all the separations between them lie in the inertial
range.  In particular the 2nd order correlation function written
explicitly is
\begin{eqnarray}
 &&{\cal F}_2^{\alpha\beta}( {\B.r}_{0}| {\B.r}_1,
 {\B.r_1}',t_1;\B.r_2,\B.r'_2,t_2) 
 \nonumber \\ && 
=\langle {\cal W}^{\alpha}( {\B.r}_0| {\B.r}_1,\B.r'_1,t_1) 
{\cal W}^{\beta}( {\B.r}_{0} | {\B.r}_2,\B.r'_2,t_2)\rangle\ .
 \label{Fab}
\end{eqnarray}
In addition to the $n$-order correlation functions the statistical
theory calls for the introduction of a similar array of response or
Green's functions.  The most familiar is the 2nd order Green's
function $G^{\alpha\beta} ({\B.r}_{0} |X_1;x_2)$ defined by the
functional derivative
 \begin{equation}
 G ^{\alpha\beta}({\B.r}_{0} |X_1;x_2) =
 \left\langle { \delta {\cal W}^{\alpha}( {\B.r}_0|X_1)  
 \over  \delta \phi^{\beta}(\B.r_0| x_2)}\right\rangle \ .
 \label{green} 
 \end{equation}
 In stationary turbulence these quantities depend on $t_1-t_2$ only,
 and we can denote this time difference as $t$.
 
 Consider next the nonlinear Green's functions $\B.{\cal G}_{m,n}$
 which are the response of the direct product of $m$ BL-velocity
 differences to $n$ perturbations. In particular \FL
 \begin{eqnarray}
 &&\B.{\cal G}_{2,1}({\B.r}_0|X_1,X_2;x_3)=
 \left<{\delta [\B.{\cal W}({\B.r}_0|X_1)
  \B.{\cal W}({\B.r}_0|X_2)]\over \delta \B.\phi({\B.r}_0|x_3)}
\right>,\nonumber \\
 &&\B.{\cal G}_{1,2}({\B.r}_0|X_1;x_2,x_3)=
 \left<{\delta^2 \B.{\cal W}({\B.r}_0|X_1)
  \over  \delta \B.\phi({\B.r}_0|x_2)\delta\B.\phi({\B.r}_0|x_3)}
\right>, \label{newG} \\
 &&\B.{\cal G}_{2,2}(\B. r_0|X_1,X_2;x_3,x_4) =
 \left<{\delta^2 [\B.{\cal W}(\B.r_0|X_1) \B.{\cal W}(\B.r_0|X_2)]
\over
   \delta \B.\phi({\B.r}_0|x_3)\delta  \B.\phi({\B.r}_0|x_4)}
 \right>, \nonumber \\
 &&\B.{\cal G}_{3,1}({\B.r}_0|X_1,X_2,X_3;x_4)
 \nonumber \\ &&=\left<{\delta [\B.{\cal W}({\B.r}_0|X_1)
  \B.{\cal W}({\B.r}_0|X_2) \B.{\cal W}({\B.r}_0|X_3)]\over
 \delta\B.\phi({\B.r}_0|x_4)}
 \right>\ . \nonumber
 \end{eqnarray}
 Note that the Green's function $\B.G$ of Eq.~(\ref{green}) is $
 \B.{\cal G}_{1,1}$ in this notation.
\narrowtext


\subsection{Hierarchies of  Equations for the Statistical Quantities}
\subsubsection{The correlation functions}
The rate of change of the $n$-order correlation functions with respect
to any of the time variables is computed as
\begin{equation}
{\partial \B.{\cal F}_n(\B.r_0|X_1,X_2,\dots , X_n) \over dt_1}=
\langle {\partial \B.{\cal W}_1\over \partial t_1}\B.{\cal W}_2\dots
\B.{\cal W}_n \rangle \ . 
\end{equation}
Using the equation of motion (\ref{newNS}) we find
\begin{eqnarray}
&&{\partial \B.{\cal F}_n(\B.r_0|X_1,\dots , X_n) \over dt_1}
+\B.{\cal D}_n(\B.r_0|X_1,\dots , X_n)
 \nonumber \\ &&=\B.{\cal J}_n(\B.r_0|X_1,\dots , X_n) 
\ , \label{balnp}\\
&&\B.{\cal D}_n(\B.r_0|X_1,\dots , X_n)
  = \langle(\hat{\cal L} \B.{\cal W})_1 \B.{\cal W}_2\dots
\dots \B.{\cal W}_n \rangle ,
  \nonumber \\
&&\B.{\cal J}_n (\B.r_0|X_1,\dots , X_n)
= \nu(\nabla_1^2\!+\!{\nabla'}_1^2)\langle
\B.{\cal W}_1\dots \B.{\cal W}_j\dots \B.{\cal W}_n \rangle . 
\nonumber
\end{eqnarray}
with $\hat\B.{\cal L}_1\equiv\hat\B.{\cal
  L}(\B.r_0,t_0|\B.r_1,\B.r'_1,t)$.  We remember that $(\hat\B.{\cal
  L} \B.{\cal W})_1$ is a nonlocal object that is quadratic in
BL-velocity differences, cf. Eq. (\ref{calL}). In writing Eq.
(\ref{balnp}) we discarded the forcing term; it was shown before that
if the forcing is limited to the largest scale $L$, its contribution
is negligible when all the separations in the correlations functions
are much smaller than $L$.

To understand the role of the various contributions in
Eq.~(\ref{balnp}) we first note that in the limit $\nu \to 0$ the term
$ \B.{\cal J}_n (\B.r_0|X_1,\dots , X_n) $ vanishes.  To see this note
that in the fully unfused case this term is bounded from above by

\begin{figure}
\epsfxsize=8.6truecm
\epsfbox{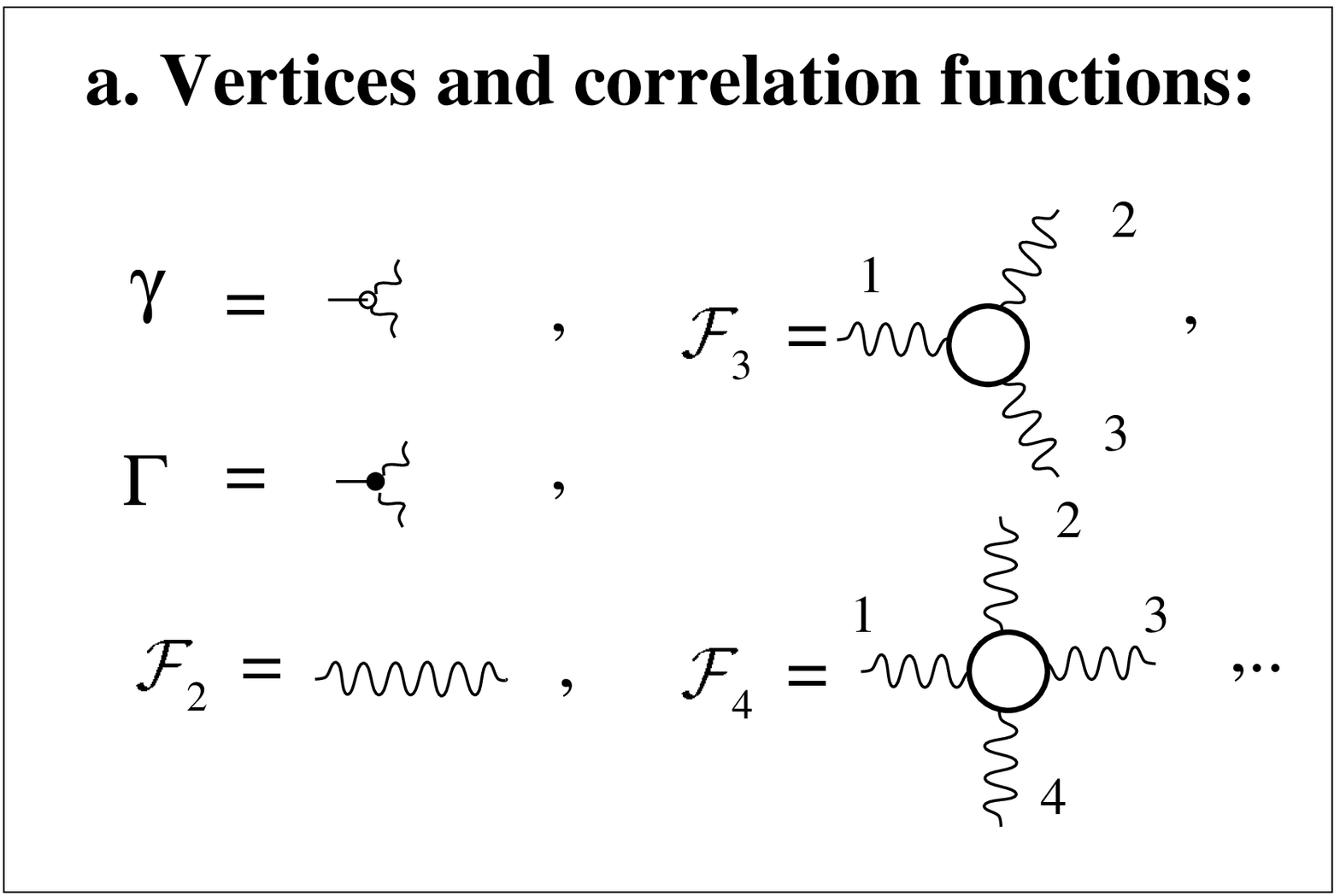}
\epsfxsize=8.6truecm
\epsfbox{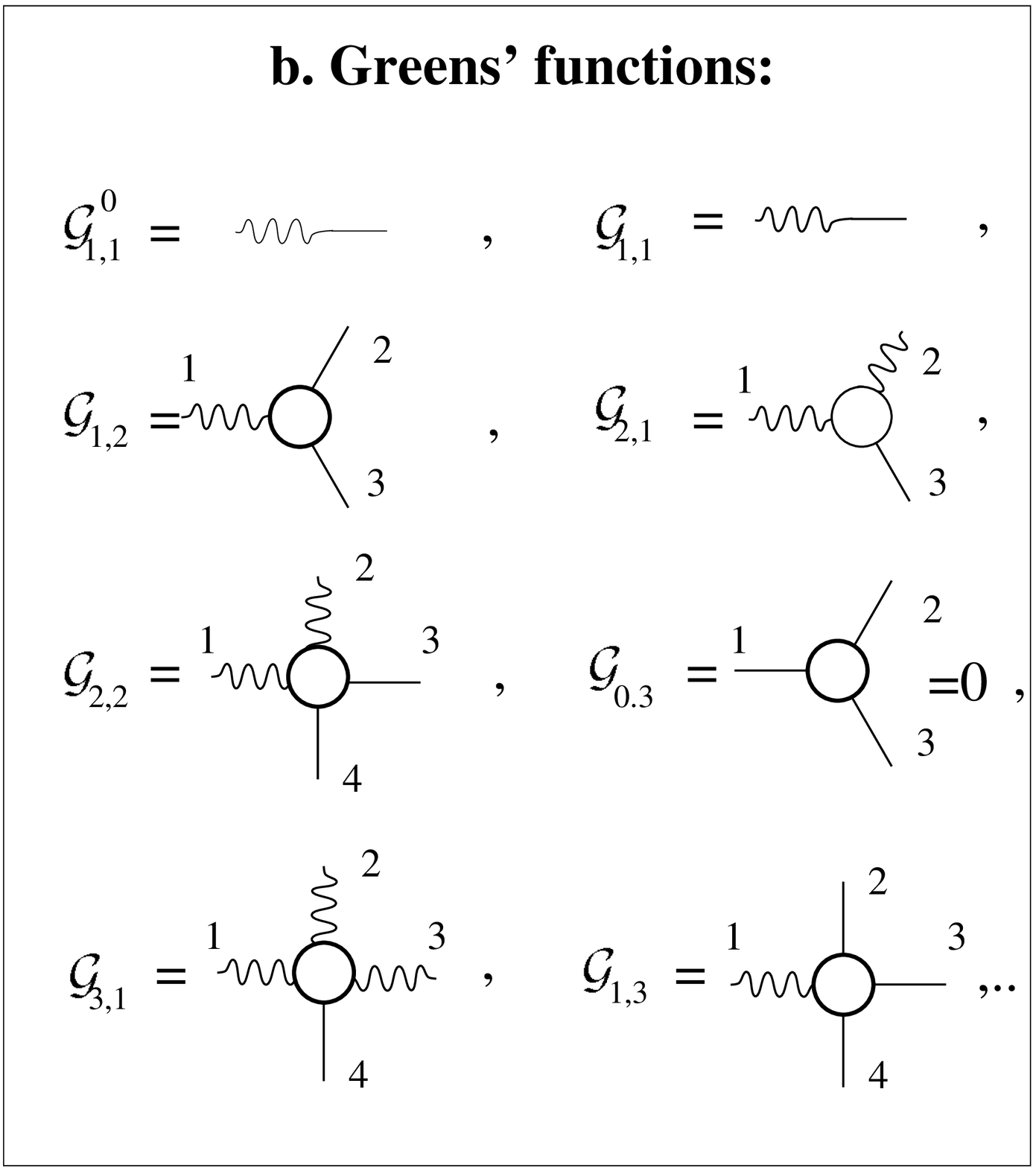}
\vskip 0.5cm 
\caption
{The diagrammatic notation of the basic objects of the theory.  Panel
  a: the vertex $\gamma$ and the correlation functions ${\cal F}_n$
  with $n=2,3,4$. The vertex $\Gamma$ has not been defined yet, but it
  appears later, cf. Eq.~(61). Panel b: the bare Green's function
  ${\cal G}^0_{1,1}$ (thin line), and the dressed Green's functions
  ${\cal G}_{n,m}$. Objects with only straight tails are identically
  zero.}
\label{F1}
\end{figure}
\noindent
$C\nu \B.{\cal J}_{n,p} (\{\B.R_j\},0)/R_{\rm min}^2$ where $C$ is
a $\nu$-independent constant and $R_{\rm min}$ is the minimal
separation between the coordinates. There is nothing in this quantity
to balance $\nu$ in the limit $\nu \to 0$. This is of course the
advantage of working with fully unfused quantities; we could not do
this with the balance equation for fused correlators [say structure
functions $S_n(R)$] as the dissipative term approaches a finite
limit when $\nu \to 0$.  Thus for $\nu\to 0$, or for very large
Reynolds numbers, we have
\begin{equation}
{\partial \B.{\cal F}_n(\B.r_0|X_1,\dots , X_n) \over dt_1}
+\B.{\cal D}_n(\B.r_0|X_1,\dots , X_n) =0\ .
 \label{balt}
\end{equation}
These equations, which are exact, can be written as a chain of
equations relating the rate of change of the $n$th order correlation
function to the ($n+1$)th order correlation function.  To this aim
introduce the vertex function
\begin{eqnarray}
&&\gamma^{\alpha\mu\sigma}(\B.r_1,\B.r'_1,
\tilde\B.r)={1\over 2}\{[P^{\alpha\mu}
(\B.r_1-\tilde \B.r)-P^{\alpha\mu}(\B.r'_1-
\tilde \B.r)]{\partial\over \partial \tilde
r_{\sigma}}\nonumber \\&&+[P^{\alpha\sigma}
(\B.r_1-\tilde \B.r)-P^{\alpha\sigma}(\B.r'_1
-\tilde \B.r)]{\partial\over \partial \tilde
r_{\mu}} \ . \label{gammanl}
\end{eqnarray}
With the help of this function we rewrite Eq.~(\ref{newNS}) in the
form \FL
\begin{eqnarray}
&&[{\partial \over \partial t_1}-\nu (\nabla_1^2+\nabla_{1'}^2)]
{\cal W}^\alpha(\B.r_0|X_1)\nonumber \\+&&\int d\tilde \B.r
\gamma^{\alpha\mu\sigma}(\B.r,\B.r',\tilde\B.r) {\cal W}
^\mu(\B.r_0|\tilde X_1)
{\cal W}^\sigma(\B.r_0|\tilde X_1)\nonumber \\
&&=\int d\tilde \B.r[ P^{\alpha\mu}(\B.r_1-\tilde \B.r)
- P^{\alpha\mu}(\B.r'_1-\tilde \B.r)]\phi^\mu
(\B.r_0|\tilde\B.r,t_1). \label{newNS2}
\end{eqnarray}
Here we used the shorthand notation 
\begin{equation}
\tilde X_1 \equiv (\tilde\B.r,\B.r_0,t_1) \ . \label{deftildx}
\end{equation}
From this we can present immediately the interaction term $\B.{\cal
  D}_n$ in the hierarchy of equations as follows
\begin{eqnarray}
&&{\cal D}^{\alpha\beta\gamma\dots}_n(\B.r_0|X_1,\dots , X_n)
=\int d \tilde \B.r\gamma^{\alpha\mu\sigma}(\B.r_1,\B.r'_1,
\tilde\B.r)\nonumber \\
&&\times{\cal F}^{\mu\sigma\beta\gamma\dots}_{n+1}(\B.r_0| \tilde X_1;
 \tilde X_1;X_2,\dots , X_{n+1}) \ . \label{baltver}
\end{eqnarray}

\begin{figure}
\epsfxsize=8.6truecm
\epsfbox{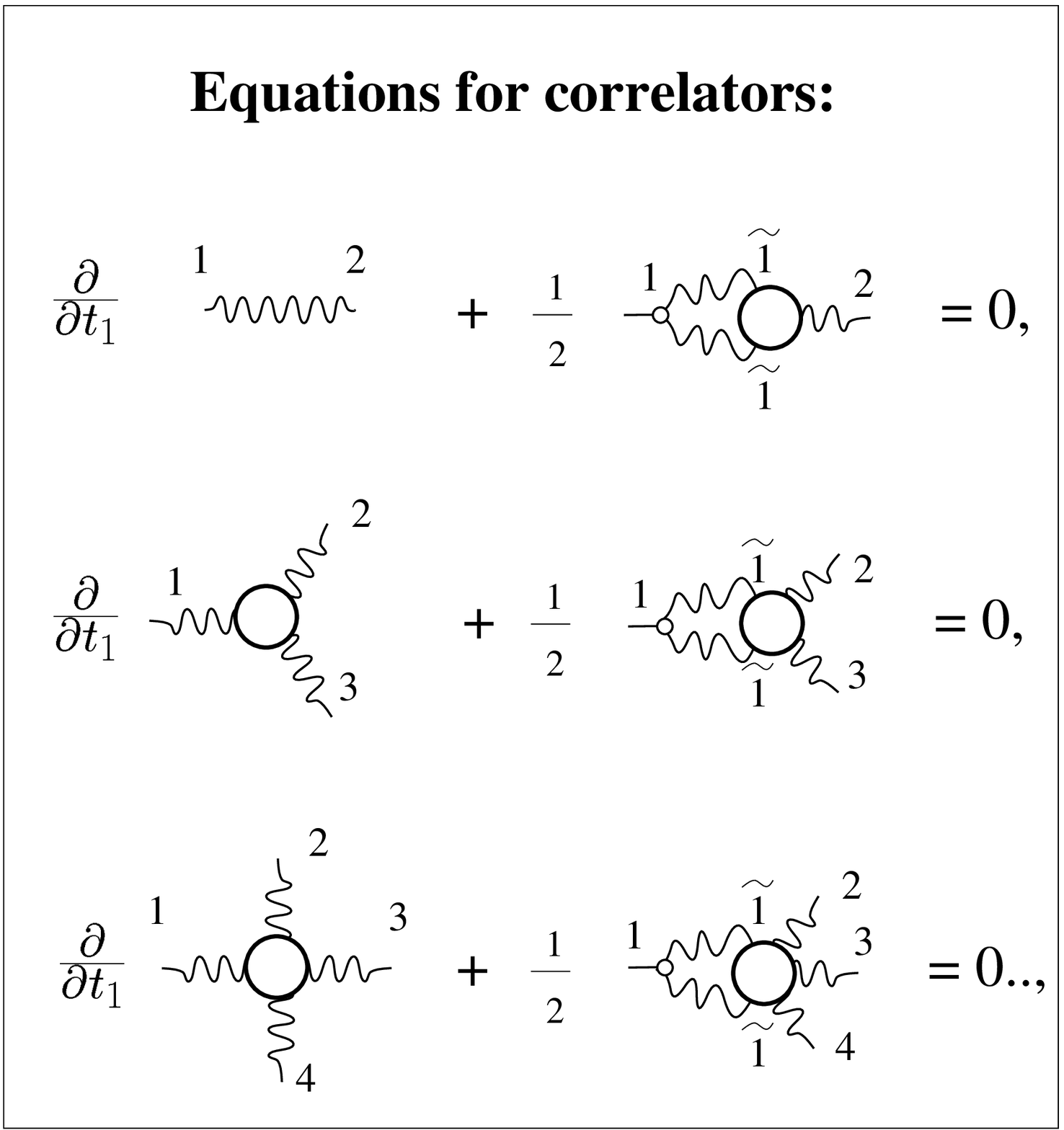}
\vskip 0.5cm 
\caption{
  The hierarchy of equation for the correlation functions.  The rate
  of change of ${\cal F}_n$ is related to an integral over ${\cal
    F}_{n+1}$ according to Eqs. (29) and (33). There are two tails
  designated by $\tilde 1$, carrying the coordinates $\tilde X_1$. An
  integration over $\tilde r$ is assumed.}
\label{F2}
\end{figure}

In order to make the structure of these equations more transparent we
will represent them graphically using the diagrammatic notations
introduced in Fig.\ref{F1}. The convention chosen here is that every
velocity difference is represented by a wavy line of unit length, and
the 2-point correlator as a wavy line of twice that length.  The
forcing is represented by a straight line of unit length. Thus the
Green's function, which is the response of the velocity field to the
forcing, is represented by a wavy line of unit length connected to a
straight line of unit length which represents the forcing. This
convention is generalized to the higher order quantities as is evident
from Fig.~\ref{F1}. The higher order quantities carry a ``junction"
that connects the appropriate number of wavy and straight lines of
unit length each. The diagrammatic representation of the hierarchy
(\ref{balt}) with the interaction term (\ref{baltver}) is shown in
Fig.~\ref{F2}.  The numbers $j=1,2,3,4$ represent sets of coordinates
$X_j$ together with the corresponding vector indices of the
correlation functions $\B.{\cal F}_n$. The two wavy lines designated
by ``$\tilde 1$" are connected to the empty circle $\B.\gamma$. This
means that both these lines carry the same set of coordinates $\tilde
X_1$, and one is supposed to integrate over $\tilde \B.r$, as is
required by Eq.~(\ref{baltver}).

\subsubsection{The Green's functions}
The equation of motion for the standard Green's function $G={\cal
  G}_{1,1}$ of Eq.~(\ref{green}) may be derived straightforwardly.
Begin with Eq.~(\ref{newNS2}) for ${\cal W}_1^\alpha$ and evaluate its
functional derivative with respect to $\phi_2^\beta$. After averaging
one gets
\begin{eqnarray}
&&\Big[{\partial \over \partial t_1}
-\nu (\nabla_1^2+\nabla_{1'}^2)\Big]{\cal G}^{\alpha\beta}_{1,1}
(\B.r_0|X_1;x_2)\nonumber\\&&+\int d\tilde \B.r\gamma
^{\alpha\mu\sigma}(\B.r_1,\B.r'_1,
\tilde\B.r) {\cal G}^{\mu\sigma\beta}_{2,1}(\B.r_0|\tilde X_1,
\tilde X_1;x_2)\nonumber \\&&=G^{(0)\alpha\beta}(\B.r_1,
\B.r'_1,\B.r_2)
\delta(t_1-t_2)\ . \label{balG1}
\end{eqnarray}
The RHS of this equation displays the (zero-time) bare Green's
function of Eq.~(30), which consists of a difference between two
transverse projection operators: \FL
\begin{equation}
G^{(0)\alpha\beta}(\B.r_1,\B.r'_1,\B.r_2)=P^{\alpha\beta}(\B.r_1-\B.r_2)-
P^{\alpha\beta}(\B.r'_1-\B.r_2). \label{G11}
\end{equation}
The nonlinear term in the equation of motion (\ref{newNS2}) results in
the appearance of the next-order Green's function $\B.{\cal G}_{2,1}$
in the equation of motion (\ref{balG1}). In its turn, the equation of
motion of $\B.{\cal G}_{2,1}$ will involve the next-order Green's
function $\B.{\cal G}_{3,1}$. We begin to build a hierarchy of
equations for $\B.{\cal G}_{n,1}$ similar to the hierarchy of
equations for the correlation functions ${\cal F}_n$.

\begin{figure}
\epsfxsize=8.6truecm
\epsfbox{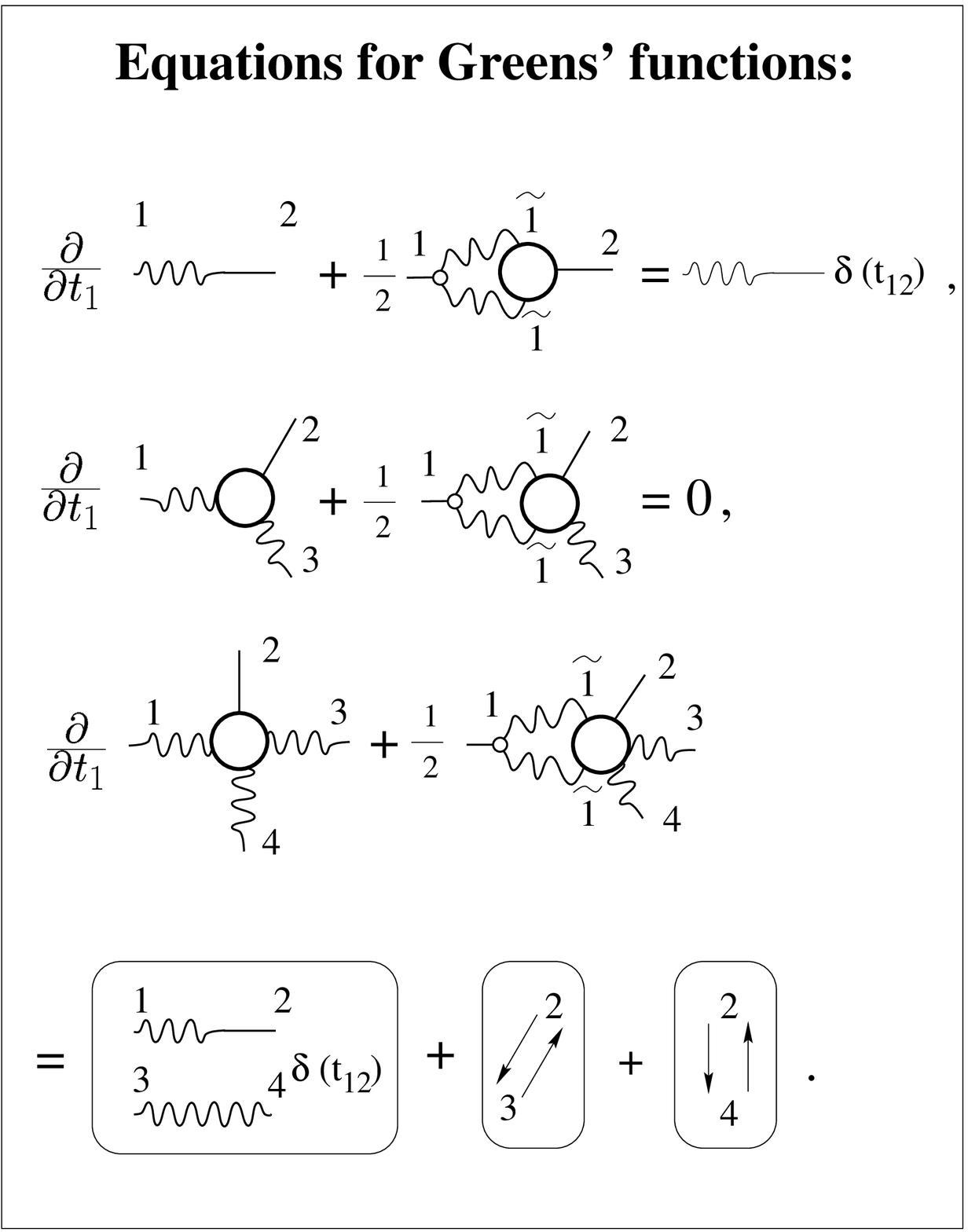}
\vskip 0.5cm 
\caption
{The hierarchy of equation for the Green's functions. The notation is
  the same as in Fig.2. The two last terms on the RHS of the last
  equation can be obtained from the first one (in the same frame) by
  permuting $2\leftrightarrow 3$ and $2\leftrightarrow 4$, as is
  shown.}
\label{F3}
\end{figure}

 To derive this hierarchy
we begin again with Eq.~(\ref{newNS2}) for ${\cal W}_1^\alpha$,
multiply the equation by ${\cal W}_2^\beta\dots{\cal W}_n^\psi$, and
then evaluate the functional derivative with respect to
$\phi^\omega_{n+1}$. After averaging and discarding the viscous term
the resulting equations read
\begin{eqnarray}
&&{\partial \over \partial t_1}{\cal G}
^{\alpha\beta\dots\psi\omega}_{n,1}
(\B.r_0|X_1,X_2,\dots,X_n;x_{n+1})
\nonumber\\&&+\int d\tilde \B.r\gamma^{\alpha\mu\sigma}
(\B.r_1,\B.r'_1, \tilde\B.r) \label{balG11}
\\&&\times{\cal G}^{\mu\sigma\beta\gamma
\dots\psi\omega}_{n+1,1}(\tilde X_1,\tilde X_1,
X_2,\dots,X_n;x_{n+1})\nonumber \\&&={\cal G}_{n,1}^{(0)
\alpha\beta\dots\omega}(\B.r_0|X_1,X_2\dots X_n,
\B.r_{n+1},t_1+0)\delta(t_1-t_{n+1})\ .\nonumber 
\end{eqnarray}
The bare Green's function of $(n,1)$ order on the RHS of this equation
are the following decomposition:
\begin{eqnarray}
&&{\cal G}_{n,1}^{(0)\alpha\beta\dots\psi\omega}
(\B.r_0|X_1,X_2\dots X_n,
\B.r_{n+1},t_1+0)\nonumber \\&&\equiv G^{(0)
\alpha\omega}(\B.r_1,\B.r'_1,\B.r_{n+1})
{\cal F}^{\beta\gamma\dots\psi}_{n-1}(\B.r_0|X_2,
\dots , X_{n-1})\ . \label{Gn1}
\end{eqnarray}
The diagrammatic representation of the first three equations in the
hierarchy for $\B.{\cal G}_{n,1}$ are shown in Fig.~\ref{F3}. The main
difference of this hierarchy from Eqs.~(\ref{balt}) is the presence of
inhomogeneous terms on the RHS. The second equation in Fig.~\ref{F3}
is atypical in having a zero on the RHS. The origin of this zero is
simply the fact that $\B.{\cal F}_1=0$.

In a similar manner we can derive additional hierarchies of equations
for $\B.{\cal G}_{n,2}$, $\B.{\cal G}_{n,3}$, etc. We do this starting
from the same equation of motion (\ref{newNS2}) for $\B.{\cal W}_1$,
multiplying it by $\B.{\cal W}_2\dots\B.{\cal W}_n$, and then
evaluating the second order, third order, etc. functional derivative
with respect to the forcing. The resulting hierarchy of equations
reads
\begin{eqnarray}
&&{\partial \over \partial t_1}\B.{\cal G}_{n,p}
(\B.r_0|X_1,X_2,\dots,X_n,x_{n+1},\dots x_{n+p})
\nonumber\\ && +\int d\tilde \B.r\B.\gamma^(\B.r_1,\B.r'_1,
\tilde\B.r) \label{balGn} \\ && \times\B.{\cal G}_{n+1,p}
(\B.r_0|\tilde X_1;
\tilde X_1,X_2,\dots,X_n;x_{n+1}\dots x_{n+p})
\nonumber \\ && =\sum_{j=1}^p\B. G^{(0)}(\B.r_0|
\B.r_1,\B.r'_1,\B.r_{n+j})\delta(t_1-t_j)
\B.{\cal G}_{n-1,p-1}(\B.r_0|X_2\dots
\nonumber \\
&&\ \ \  \dots X_n,
\B.r_{n+1},\dots, \B.r_{n+j-1},\B.r_{n+j+1},
\dots \B.r_{n+p})\ . \nonumber
\end{eqnarray}
In this equation the viscous terms have again been discarded; they are
vanishingly small in the limit $\nu\to 0$ compared to the other terms
in the equations for the fully unfused Green's functions.
\section{The temporal dependence of the statistical 
quantities: Temporal Multiscaling}

In this section we discuss the temporal dependence of our space-time
correlation functions, culminating with a new representation for
them, see Eq.~(\ref{repres}) below. This representation will be very
useful in setting up our scheme to compute the scaling exponents.
Along the way we will prove the statement that if the strong
assumption presented in Eq.~(\ref{tscaleinv}) were true, than our
hierarchy of equations would have predicted a linear dependence of
$\zeta_n$ on $n$. Our first step then is to show that
Eq.~(\ref{tscaleinv}) is in contradiction with a nonlinear dependences
of $\zeta_n$ on $n$.
\subsection{Locality of the Interaction Integrals}
In this subsection we discuss the locality of the integrals appearing
in the hierarchic equations (\ref{balt}) and (\ref{baltver}), where
``locality" here means that the integrals converge in the infrared
and in the ultraviolet limits. This property is important, since it
excludes the appearance of a renormalization scale as a cutoff on the
spatial integrals. Moreover, we will show that the locality of the
integrals in addition to the assumption of full scale invariance (7)
imply linear scaling. Thus the only way to get anomalous scaling is to
show that (7) is incorrect, as we will do in the next section.
Fortunately, the property of locality of the integral appearing in
(\ref{baltver}) was established already in previous publications. It
stems from the existence of the fusion rules which govern the
asymptotic properties of the correlation functions when a group of
coordinates coalesce together \cite{LPfuse,97LP}. To control the
ultraviolet properties of the integrals we need to analyze the
situation when the dummy integration coordinate $\tilde\B.r$
approaches either $\B.r_0$, or any of the other coordinates $\B.r_i$.
The asymptotic behavior of the integrand is then determined by the
fusion rules for the fusion of two coordinates while all the rest of
the coordinates remain separated by much larger distances. The
infrared limit is obtained when $\tilde\B.r\to \infty$, and the
asymptotics of the integrand are governed by fusion rules for $n-1$
coordinates coalescing together compared to a single coordinate that
approaches an infinite distance from all the rest.  Using the fusion
rules in this way one can prove convergence in a wide
window of the numerical values of the scaling exponents around their
K41 or experimental values \cite{97LP}.
\subsection{The Consequences of Locality}
The differential equations (\ref{balt}) can be turned readily into
integral equations by integrating over the time variable.  We argued
above that the spatial integration converges in the infrared
{\em and} the ultraviolet regimes.  Here we explain the consequences
of this property of locality.  The needed material was developed in
some detail in Ref.\cite{97LPP} but we repeat here all the essentials.

In general our $n$-order correlation functions may depend on $n$
different times. If Eq.~(7) were correct, the correlation functions
would be a homogeneous function of all the time coordinates. On the
other hand, some of these times could be the same, and when they are
all the same we get our simultaneous object (\ref{defF}) which is
believed to be homogeneous in its spatial coordinates.  The simplest
many-time case is the one in which there are two different times in
(\ref{defFtime}).  Chose $t_i=t+s$ for every $i\le p$ and $t_i=t$ for
every $i>p$. We will denote the correlation function with this choice
of times as $\B.{\cal F}_{n,1}^{(p)}(s)$, omitting for brevity the
rest of the arguments. Introduce the typical decorrelation time
$s_{n,1}^{(p)}(R)$ that is associated with the one-time difference
quantity $\B.{\cal F}^{(p)}_{n,1}(s)$ when all the spatial separations
are of the order of $R$:
\begin{equation}
\int^{\infty}_0 ds \B.{\cal F}^{(p)}_{n,1}(s) \equiv
s_{n,1}^{(p)} \B.{\cal F}^{(p)}_{n,1}(0) \ . \label{deftau1}
\end{equation}
In Ref.\cite{97LPP} we showed that the consequence of locality is that
$s_{n,1}^{(p)}$ satisfies a $p$-independent scaling law:
\begin{equation}
s_{n,1}(R) \sim R S_{n-1}(R)/ S_n(R)\propto  R^{z_{n,1}}\ .
 \label{tau1}
\end{equation}
There we introduced the dynamical scaling exponent $z_{n,1}$ that
characterizes this time and found that
\begin{equation}
z_{n,1} = 1+\zeta_{n-1}-\zeta_n \ . \label{zn1}
\end{equation}

Now we ask whether the same time scale also characterizes correlation
functions having two or more time separations. Consider
the three-time quantity that is obtained from $\B.{\cal F}_n$ by
choosing $t_i=t+s_1$ for $i\le p$, $t_i=t+s_2$ for $p<i\le p+q$, and
$t_i=t$ for $i>p+q$. We denote this quantity as $\B.{\cal
  F}_{n,2}^{(p,q)}(s_1,s_2)$, omitting again the rest of the
arguments.  We define the decorrelation time $s_{n,2}^{(p,q)}$ of this
quantity by
\begin{equation}
\int^{\infty}_0 ds_1 ds_2 \B.{\cal F}_{n,2}^{(p,q)}
(s_1,s_2)
\equiv
[s_{n,2}^{(p,q)}]^2 \B.{\cal F}_{n,2}^{(p,q)}(0,0)\ .
 \label{deftau2}
\end{equation}
One could think naively that the decorrelation time
$s_{n,2}^{(p,q)}$ is of the same order as (\ref{tau1}). The
calculation \cite{97LPP} leads to a different result:
\begin{equation}
[s_{n,2}(R)]^2 \sim R^2 S_{n-2}(R)/ S_n(R)\propto [R^{z_{n,2}}]^2
\ . \label{tau2}
\end{equation}
We see that the naive expectation implied by
Eq.~(\ref{tscaleinv}) is not realized.  The scaling exponent of the
present time is different from (\ref{zn1}):
\begin{equation}
z_{n,2} =1+(\zeta_{n-2}-\zeta_n)/2 \ . \label{zn2}
\end{equation}
The difference between the two scaling exponents $z_{n,1}-z_{n,2}=
\zeta_{n-1}-(\zeta_n+\zeta_{n-2})/2$. This difference is zero for
linear scaling, meaning that in such a case the naive expectation that
the time scales are identical is correct. On the other hand for the
situation of multiscaling the Hoelder inequalities require the
difference to be positive. Accordingly, for $R\ll L$ we have
$s_{n,2}(R)\gg s_{n,1}(R)$.

We can proceed to correlation functions that depend on $m$ time
differences. Omitting the upper indices which are irrelevant for the
scaling exponents we denoted \cite{97LPP} the correlation function as
$\B.{\cal F}_{n,m}(s_1\dots s_m)$, and established the exact scaling
law for its decorrelation time.  The definition of the decorrelation
time is
\begin{equation}
\int^{\infty}_0 ds_1\dots ds_m\B.{\cal F}_{n,m}
(s_1\dots s_m)
\equiv [s_{n,m}]^m \B.{\cal F}_{n,m}(0\dots 0).\label{gentau}
\end{equation}
We found the dynamical scaling exponent that characterizes
$s_{n,m}$ when all the separations are of the order of $R$,
$s_{n.m}\propto R^{z_{n,m}}$:
\begin{equation}
z_{n,m} =1 +(\zeta_{n-m}-\zeta_n)/m \ , \quad n-m\le 2 . 
\label{znm}
\end{equation}
One can see, using the Hoelder inequalities, that $z_{n,m}$ is a
nonincreasing function of $m$ for fixed $n$, and in a multiscaling
situation they are decreasing. The meaning is that the larger is $m$,
the {\em longer} is the decorrelation time of the corresponding
$m+1$-time correlation function, $ s_{n,p}(R)\gg s_{n,q}(R)\ \ {\rm
  for}\ \ p<q\ $.

It is obvious now that the assumption of complete scale invariance
(\ref{tscaleinv}), which is tantamount to the assertion that $z_{n,m}$
is $m$-independent, necessarily requires a linear dependence of
$\zeta_n$ on $n$. An $m$-independent $z_{n,m}$ means that
\begin{equation}
\zeta_{n-m}-\zeta_n = 
{\rm const}\times m \ , \quad{\rm linear~scaling!} \ .
\end{equation}
The results (\ref{znm}) shows that in a multi-scaling situation our
correlation functions cannot exhibit scale invariance.

To expose the consequences in a complementary way that will be useful
in what follows, we considered higher order temporal moments of the two-time
correlation functions:
\begin{equation}
\int^{\infty}_0 ds s^{k-1}
 \B.{\cal F}^{(p)}_{n,1}(s) \equiv
({\overline{s^k}})^{(p)}_{n,1}
 \B.{\cal F}^{(p)}_{n,1}(0)\ .
\label{deftauk}
\end{equation}
The intuitive meaning of $({\overline{s^k}})^{(p)}_{n,1}$ is a {\em
  $k$-order decorrelation moment} of $\B.{\cal F}^{(p)}_{n,1} (R,s)$
whose dimension is (time)$^k$.  The first order decorrelation moment
is the previously defined decorrelation time $s_{n,1}^{(p)}$. We found
the scaling laws
\begin{equation}
 [\overline{s^k}]_{n,1}^{(p)}\sim  (s_{n,k})^k \sim 
R^k S_{n-k}(R)/  S_n(R) \label{relk}
\end{equation}
for $k \le n-2 $. The procedure does not yield information about
higher $k$ values. We learn from the analysis of the moments that
there is no single typical time which characterizes the $s$ dependence
of $\B.{\cal F}^{(p)}_{n,1}$.  There is no simple ``dynamical scaling
exponent" $z$ that can be used to collapse the time dependence in the
form $\B.{\cal F}^{(p)}_{n,1}(s)\sim R^{\zeta_n}f(s/R^z)$.  Even the
two-time correlation function is not a scale invariant object. In this
respect it is similar to the probability distribution function of the
velocity differences across a scale $R$, for which the spectrum of
exponents $\zeta_n$ is a reflection of the lack of scale invariance.
\subsection{Temporal Multiscaling Representation}
This subsection offers a convenient presentation of the time
dependence of the correlation functions.  Consider first the
simultaneous function $\B.{\cal T}_n(\B.r_1\dots \B.r'_n)$.  Following
the standard ideas of multifractals \cite{Fri,86HJKPS} the
simultaneous function can be represented as
\begin{eqnarray}
&&\B.T_n(\B.r_1,\B.r'_1\dots \B.r'_n)= U^n\int_{h_{\rm min}}
^{h_{\rm max}}
d\mu(h)\left({R_n\over L}\right)^{nh+{\cal Z}(h)}\nonumber \\ &&
\times \tilde T_{n,h}(\B.\rho_1,\B.\rho'_1,\dots,\B.\rho'_n)
\ . \label{repres0}
\end{eqnarray}
Here $U$ is a typical velocity scale, and we have introduced the ``typical scale
of separation" of the set of coordinates
\begin{equation}
R^2_n={1\over n}\sum_{j=1}^n |\B.r_j-\B.r_j^{\prime}|^2 \ . \label{radgyr}
\end{equation}
Greek coordinates stand for dimensionless (rescaled) coordinates, i.e.
\begin{equation}
\B.\rho_j=\B.r_j/R_n \ , \quad \B.\rho'_j=\B.r'_j/R_n \ , \
\end{equation}
and finally the function ${\cal Z}(h)$ is defined as
\begin{equation}
{\cal Z}(h) \equiv 3- {\cal D}(h)\ . \
\end{equation}
The function ${\cal Z}(h)$ is related to the scaling exponents
$\zeta_n$ via the saddle point requirement
\begin{equation}
\zeta_n = \min_h[nh+ {\cal Z}(h)] \ . \
\end{equation}  
This identification stems from the fact that the integral in
(\ref{repres0}) is computed in the limit $R_n/L\to 0$ via the steepest
descent method. Neglecting logarithmic corrections one finds that
$\B.{\cal F}_n\propto R_n^{\zeta_n}$.

The physical intuition behind the representation (\ref{repres0}) is
that there are velocity field configurations that are characterized by
different scaling exponents $h$. For different orders $n$ the main
contribution comes from that value of $h$ that determines the position
of the saddle point in the integral (\ref{repres0}).

There is no reason why not to extend this intuition to the time
domain. The particular velocity configurations that are characterized
by an exponent $h$ also display a typical time scale $t_{R,h}$ which
is written as
\begin{equation}
t_{R,h} \sim {R\over U} \left({L\over R}\right)^h \ . 
\end{equation} 
Accordingly we propose a new temporal multiscaling representation for
the time dependent function
\begin{eqnarray}
&&\B.{\cal F}_n(\B.r_0|X_1,\dots ,X_n)= U^n
\int\limits_{h_{\rm min}}^{h_{\rm max}}
d\mu(h)\left({R_n\over L}\right)^{nh+{\cal Z}(h)}\nonumber\\ &&
\times \tilde\B.{\cal F}_{n,h}(\B.r_0|\Xi_1,\Xi_2,\dots \Xi_n) \ . 
\label{repres}
\end{eqnarray}
The function $\tilde\B.{\cal F}_{n,h}$ depends on dimensionless
(rescaled) coordinates
\begin{equation}
\Xi_j \equiv (\B.\rho_j, \B.\rho'_j,\tau_j)\ ,
 \quad \tau_j=t_j/t_{R_n,h} \ . 
\end{equation}
Of course, we require that the function $\tilde\B.{\cal
  F}_{n,h}(\B.r_0|\Xi_1, \dots \Xi_n)$ is identical to $\tilde
T_{n,h}(\B.\rho_1,\B.\rho'_1,\dots,\B.\rho'_n)$ when its rescaled time
arguments are all the same.

The point of this presentation is that it reproduces all the scaling
laws that are involved in time integrations and differentiations. For
example, specializing to the case of one time difference as discussed
in Section 4, we can perform the integral appearing in
Eq.~(\ref{deftauk}), and derive immediately the result in
(\ref{relk}).
\section{Setting up a calculation of the scaling exponents}
This section contains the main result of this paper, i.e. the
equations that we propose as the starting point for the calculation of
the scaling exponents.

One of the main insights gained by understanding the temporal
properties of the correlation functions is that it is not true that
the hierarchy of equations for the correlation functions dictate
classical scaling by power counting. The way to see this is to
substitute the temporal multiscaling representation (\ref{repres}) in
the hierarchy of equations (\ref{baltver}), and see that power
counting gives no information: \FL
\begin{eqnarray}
&&\int_{h_{\rm min}}^{h_{\rm max}}
d\mu(h)\left({R_n\over L}\right)^{(n+1)h
+{\cal Z}(h)}\Big\{{\partial\over \partial 
\tau_1}\tilde \B.{\cal F}
_{n,h}(\Xi_1,\dots,\Xi_n)\nonumber \\&&+\int 
d\tilde\B.\rho \B.\gamma(\B.\rho_1,\B.\rho_2,\tilde\B.\rho)
\left({R_{n+1}\over R_n}\right)^{(n+1)h+{\cal Z}(h)} 
\label{great} \\ &&\times 
\tilde\B.{\cal F}_{n+1,h}(\tilde \Xi, \tilde \Xi,\Xi_2,
\dots \Xi_n)\Big\}=0 \ .\nonumber
\end{eqnarray}
In this equation $R_{n+1}$ is the mean scale of separation of the
$2(n+1)$ space coordinates of the function $\B.{\cal F}_{n+1}$
similarly to (\ref{radgyr}). This set of equations can be considered
as a linear set of functional equations for the structure functions in
rescaled coordinates. Since this equation has to be valid for any
value of $L$, and since we integrate over a positive measure, the
equation is satisfied only if the terms in curly parentheses vanish.
In other words, we will seek solutions to the equations
\begin{eqnarray}
&&{\partial\over \partial 
\tau_1}\tilde \B.{\cal F}
_{n,h}(\Xi_1,\dots,\Xi_n) +\int d\tilde\B.\rho \B.
\gamma(\B.\rho_1,\B.\rho_2,\tilde\B.\rho)
\label{greater} \\ &&\times \left({R_{n+1}\over R_n}
\right)^{(n+1)h+{\cal Z}(h)}
\tilde\B.{\cal F}_{n+1,h}(\tilde \Xi, \tilde \Xi,\Xi_2,
\dots \Xi_n)=0 \ . \nonumber 
\end{eqnarray}
We note the important fact that this equation is
invariant to the rescaling
\begin{equation}
\B.r_i\to \lambda \B.r_1 \ , \quad t_i\to \lambda^{1-h}t_i \ ,
\end{equation}
for any value of $\lambda$ and $h$. Accordingly, power counting leads
to no result. As a consequence the information about the scaling
exponents $\zeta_n$ is obtainable only from the solvability conditions
of this equation. In other words, the information is buried in
coefficients rather than in power counting. The spatial derivative in
the vertex on the RHS brings down the unknown function ${\cal Z}(h)$,
and its calculation will be an integral part of the computation of the
exponents.

We recognize nevertheless that this set of equations forms an infinite
hierarchy. To proceed we need to find intelligent ways to truncate the
hierarchy without reintroducing power counting. To this aim we are
going to use renormalized perturbation theory, which is one of the
best schemes available to express higher order quantities in terms of
lower order ones. We review the needed material in the next section.
\begin{figure}
\epsfxsize=8.6truecm
\epsfbox{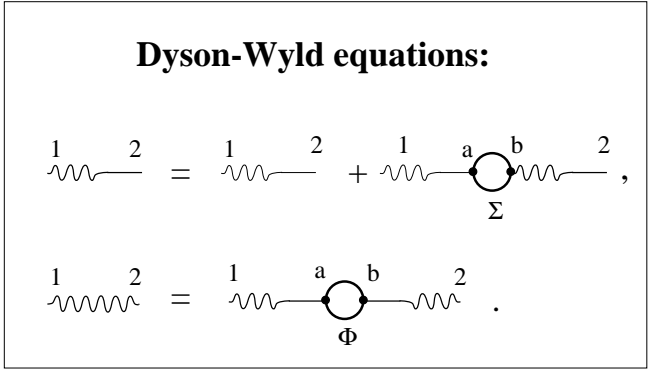}
\vskip 0.5cm 
\caption
{The coupled Dyson (first line) and Wyld (second line) equations for
  the Green's functions and second order correlator respectively. The
  mass operators are displayed in Fig.~5}
\label{F4}
\end{figure}

\section{Resummations in successive orders}
In this section we discuss successive resummations of renormalized
perturbation theory.  In doing so we will demonstrate that the theory
generates infinitely many
renormalized objects whose scaling properties are non-trivial. The
first step is the standard line-resummation that produces renormalized
two-point functions.  Further resummations produce three-, four-, and
higher point renormalized objects. We prove that this procedure
generates equations that in their fully resummed form are identical in
content to the exact hierarchies of equations derived above directly
from the fluid equations of motion. Next we explain how partially
resummed versions can be used to offer controlled approximations to
the full calculation.

Equation (\ref{newNS2}) can be used as a starting point for the
development of a line-renormalized perturbation theory for the
statistical quantities. The reader should note that this equation is
slightly different than the one used previously \cite{P-1} for the
same purpose.  The difference is that previously one used an equation
for a quantity that in the present notation reads $\B.{\cal
  W}(\B.r_0|\B.r,\B.r_0,t)$. Accordingly, one cannot read blindly the
results of the previous analysis. Nevertheless the differences are not
serious, and the spirit is the same as before.

The diagrammatic representation of the Dyson-Wyld equations for the
Green's function $\B.{\cal G}_{1,1}$ and the second order correlation
function $\B.{\cal F}_2$ is identical to its previous counterpart
\cite{P-1}, and shown in Fig.~\ref{F4}. In symbols the Dyson
equation reads now

\begin{figure}
\epsfxsize=8.6truecm
\epsfbox{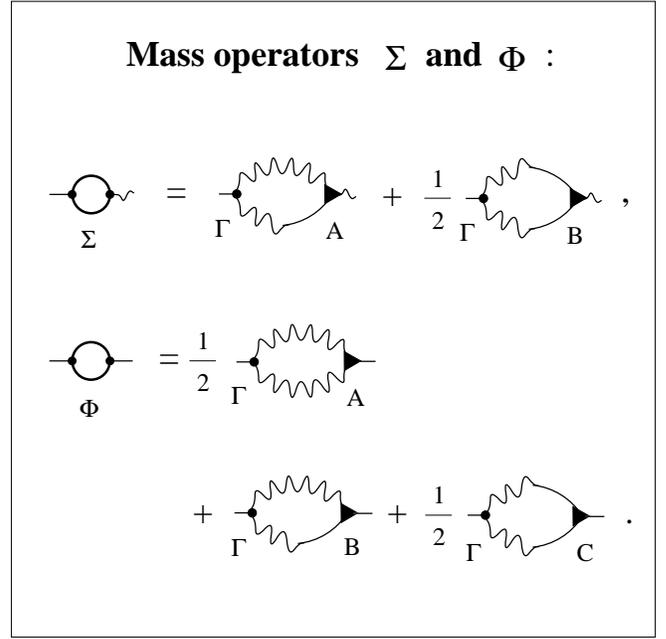}
\vskip 0.5cm 
\caption
{Exact representations of the mass operators $\Sigma$ and
  $\Phi$ in terms of the bare vertex $\Gamma$, the Green's function
  and second order correlator, and the dressed three-legged vertices
  A,B,C. }
\label{F5}
\end{figure}~~
\begin{eqnarray}
 &&{\cal G}^{\alpha\beta}_{1,1} ({\B.r}_{0}|X_1;x_2) =
 {\cal G}^{(0)\alpha\beta}_{1,1}
 ( {\B.r}_0 | X_1;x_2)   
 \nonumber \\
 &&+\int d {\B.r}_a \int_{t_2}^{t_1} dt_a {\cal G}^{(0)
\alpha\delta}_{1,1}
 ({\B.r}_0 |X_1;x_a)  \label{Dys}\\ 
&&\times\int d {\B.r}_b \int_ {t_2}^{t_a} dt_b
 \Sigma^{\delta\gamma}( {\B.r}_{0}| {\B.r}_a, {\B.r}_b ,t_a-t_b)
 {\cal G} ^{\gamma\beta}_{1,1}({\B.r}_{0}|\tilde X_b,x_2) \ ,
  \nonumber 
 \end{eqnarray}
where $\tilde X_b=\{\B.r_b,\B.r_0,t_b\}$. One sees that the number 1
 designates a set of coordinates $X_1$ and a corresponding vector
 index $\alpha$, the number 2 corresponds to $X_2$ and a corresponding
 vector index $\beta$. The meaning of ``a" is evident from the
 comparison of Eq.~(\ref{Dys}) with the first line of Fig.~\ref{F4}.
 The Wyld equation for the 2nd order correlation
function has the form
 \begin{eqnarray}
 &&{\cal F}^{\alpha\beta}_2(B.r_0 |X_1,X_2) 
 = \int d {\B.r}_a d {\B.r}_b 
 \int_{t_1}^\infty dt_a \int_{t_2}^\infty dt _b   
 \nonumber\\
 &\times& {\cal G}^{\alpha\delta}_{1,1}
 ( {\B.r}_{0} | X_1;x_a){\cal G}^{\beta\mu}(
 {\B.r}_0 |X_2;x_b)\Phi^{\delta\mu}( {\B.r}_{0}| x_a,x_b) \ .
  \label{Wyl}   
 \end{eqnarray}
 The meaning of $1,2,a$ and $b$ is again evident after comparing the
 second line of Fig.~\ref{F4} with Eq.~(\ref{Wyl}).  In equation
 (\ref{Dys}) the ``mass operator" $\Sigma$ is related to the ``eddy
viscosity" whereas in Eq.~(\ref{Wyl}) the ``mass operator" $\Phi$ is
 the renormalized ``nonlinear" noise which arises due to turbulent
 excitations. Both these quantities are given as infinite series in
 terms of the Green's function and the correlator, and thus the
 equations are coupled. The diagrammatic notation for the vertex
 $\Gamma$ is presented in Fig.~\ref{F1}a. Analytically
 $\Gamma_{\alpha\beta\gamma}(\B.r)$ is a local differential operator
 of the Euler equation

\begin{figure}
\epsfxsize=8.6truecm
\epsfbox{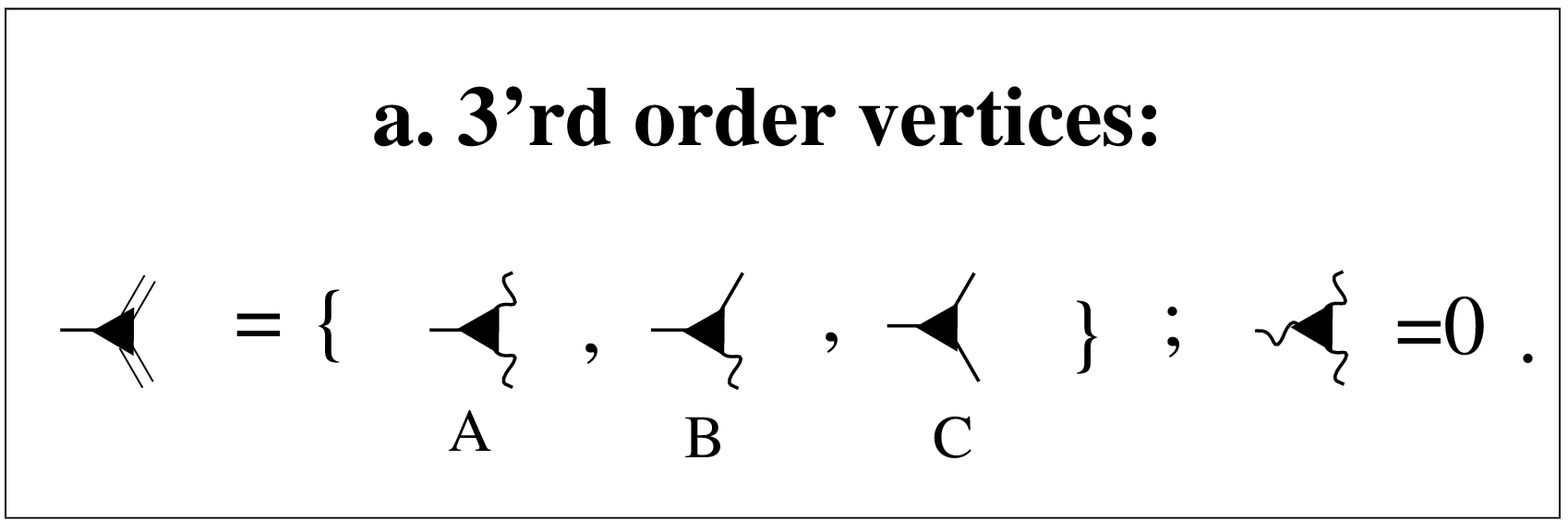}
\epsfxsize=8.6truecm
\epsfbox{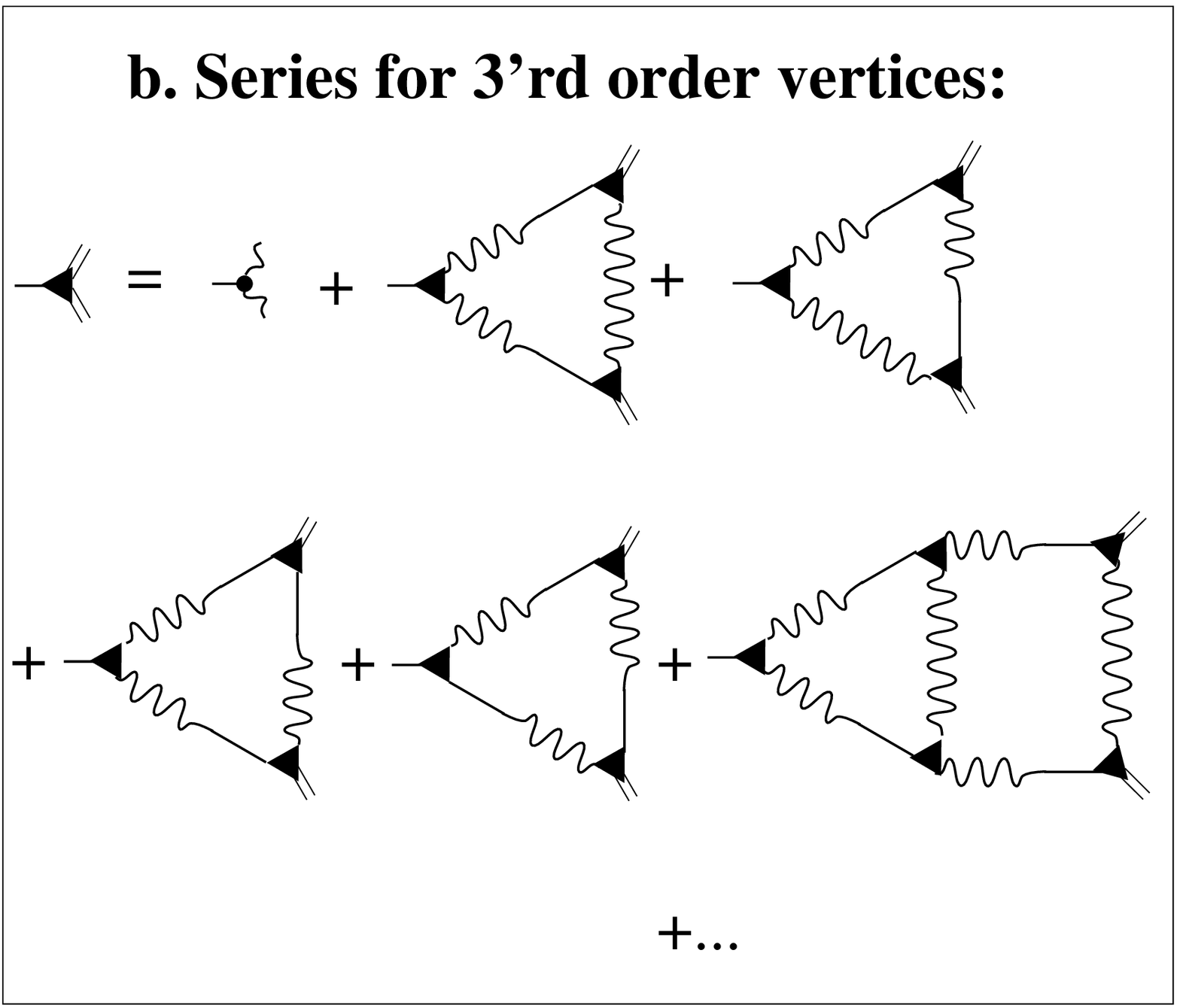}
\epsfxsize=8.6truecm
\epsfbox{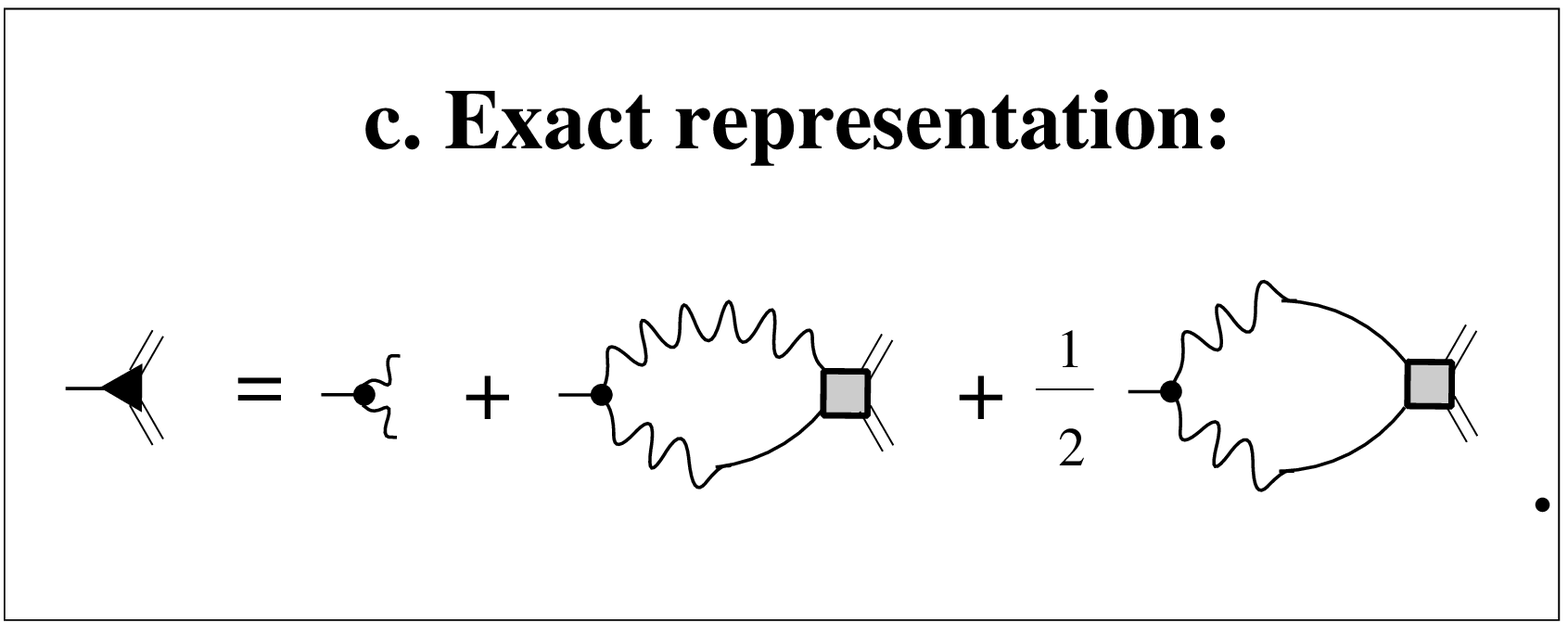} 
\vskip 0.5cm 
\caption
{The 3rd order vertices. Panel a: The three different types of 3rd
  order vertices, A, B, and C.  A double lined tail stands for either
  wavy or straight tail. There is no vertex with three wavy tails in
  the theory.  Panel b: The diagrammatic
  representation of a dressed 3rd order vertex (A, B or C) as an
  infinite series in terms of dressed 3rd order vertices and
  propagators. There is only one type of bare vertex appearing in the 
series for the dressed vertex A.  Panel c: The exact
  representation of 3rd order vertices in terms of the partly reducible
 4th order vertex (gray
  square) which is defined in Fig.~7.}
\label{F6}
\end{figure}
\vskip .3cm
\begin{figure}
\epsfxsize=8.6truecm
\epsfbox{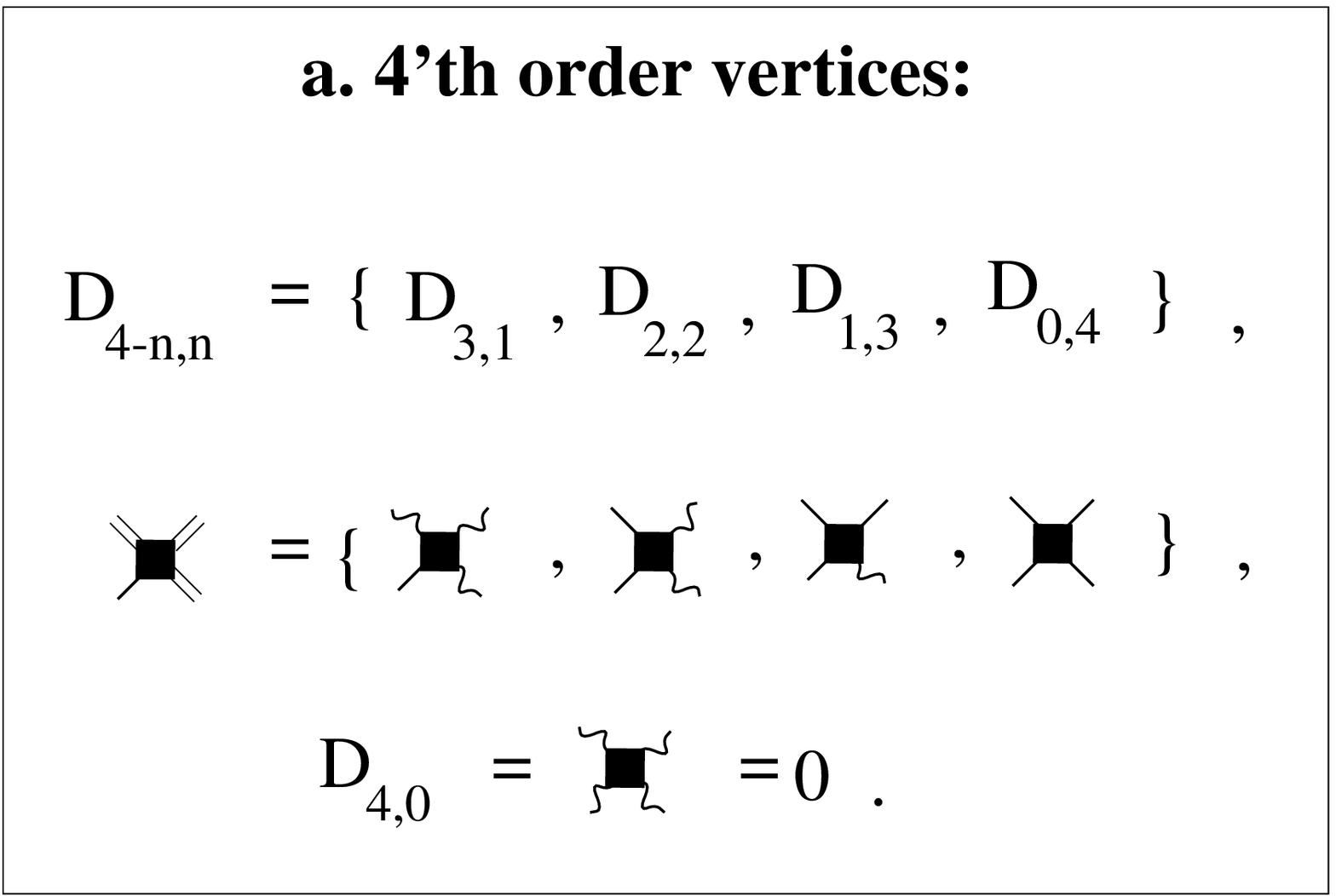}
\epsfxsize=8.6truecm
\epsfbox{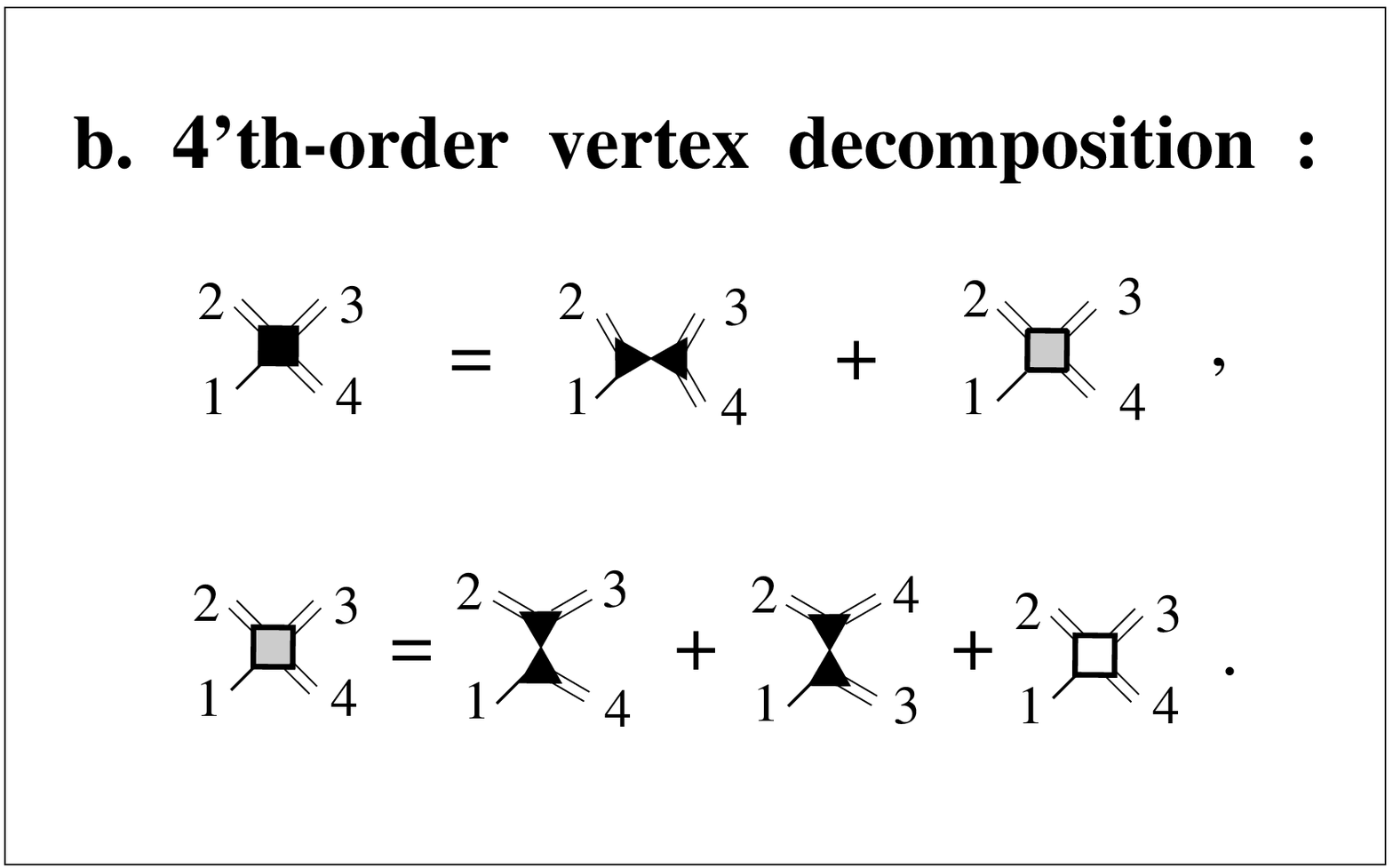}
\epsfxsize=8.6truecm
\epsfbox{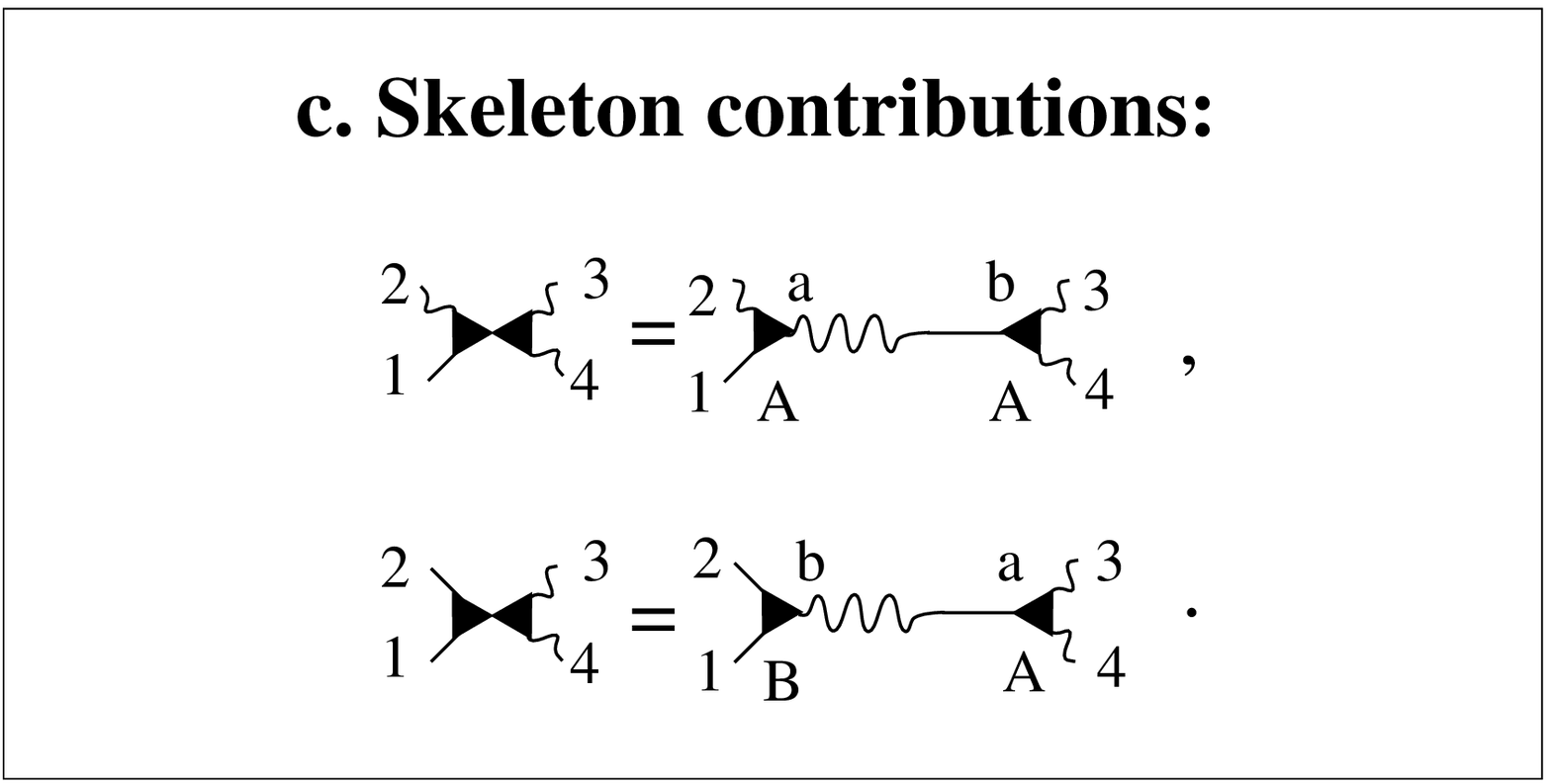}
\epsfxsize=8.6truecm
\epsfbox{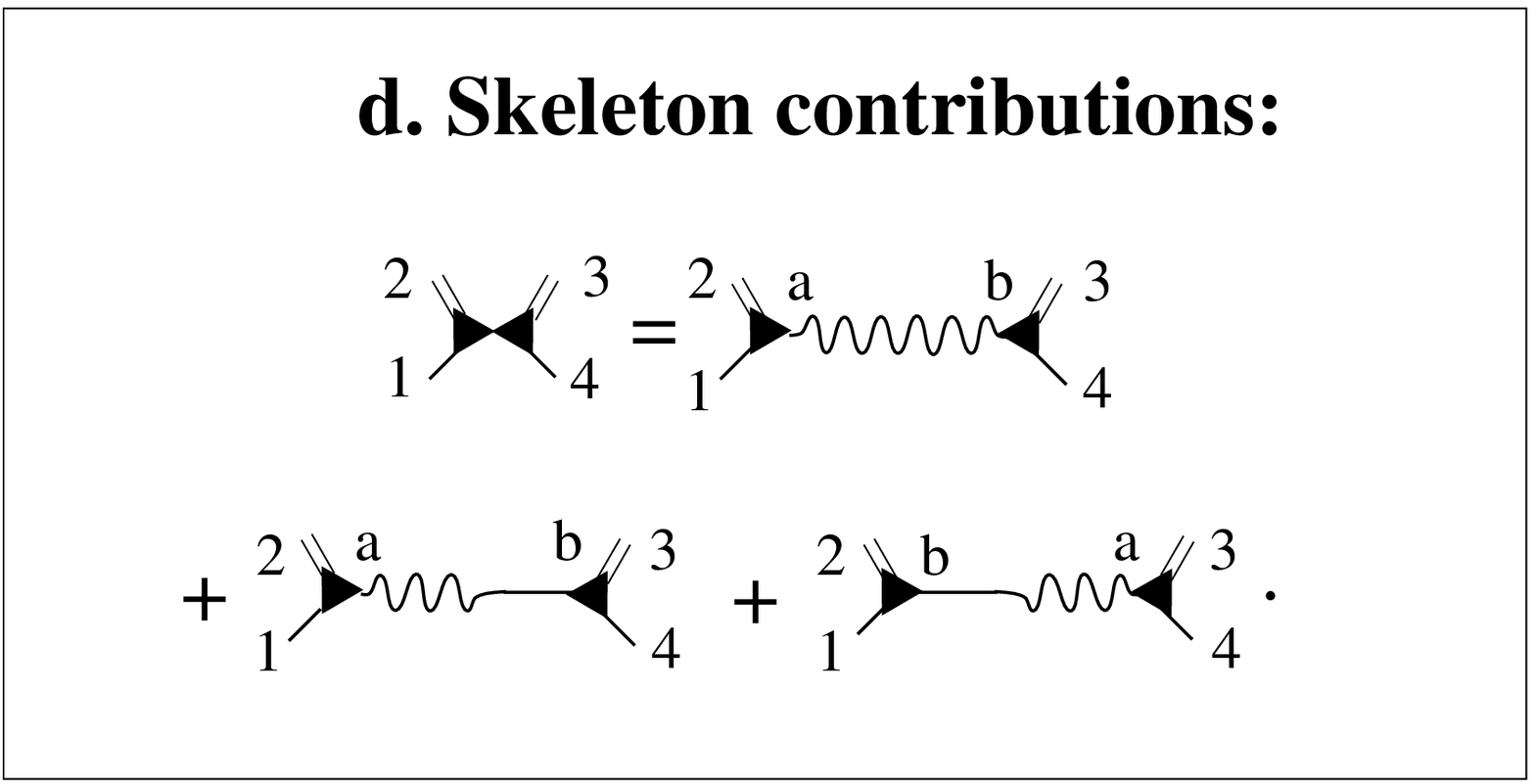}
\vskip 0.5cm 
\caption{The 4th order vertices. Panel a: the different types of 4th
  order vertices. A double lined tail stands for either wavy or
  straight tail, see text.  Panel b, first line: decomposition of 4th
  order vertices in terms of partly reducible contributions (gray
  square).  Panel b, second line: further decomposition of the vertex
  into two-eddy reducible parts and fully irreducible 4the order
  vertices (emplty square). The latter are shown in Fig.~13. Panels c
  and d: Skeleton contributions to the 4th order vertices. There are
  three terms in Panel d with three types of propagators. There is
  just one term in Panel c, the two other type of propagators require
  a tripple vertex with three wavy tails which is zero.}
\label{F7}
\end{figure}~~
\vskip .3cm
\begin{equation}
\Gamma^{\alpha\beta\gamma}(\B.r)
=-(\delta_{\alpha\gamma}\nabla_\beta
+\delta_{\alpha\beta}\nabla_\gamma)  \ . \label{barev}
\end{equation}
which is different from the non-local bare vertex $\B.\gamma$,
Eq.~\ref{gammanl}. This vertex is related to $\B.\Gamma$ via the bare
Green's function (\ref{G11}) where $G^{(0)\alpha\beta}$ is given by
Eq.~(\ref{G11}):
\begin{equation}
\gamma^{\alpha\beta\gamma}(\B.r_1,\B.r'_1,\B.r_2)= -
G^{(0)\alpha\delta}({\B.r}_1, \B.r'_1,{\B.r}_2)
\Gamma^{\delta\beta\gamma}(\B.r_2)\ , \label{gG}
\end{equation}
where only the repeated tensor index $\delta$ is summed upon.  The
bare Green's function in the BL-representation satisfies the equation
\begin{eqnarray}
 \Big[{\partial\over \partial t_1}
 &-&\nu\big(\nabla^2_1+\nabla^2_{1'}\big)\Big]
 {\cal G}_{1,1}^{(0)\alpha\beta}({\B.r}_0|X_1,x_2)
 \label{b18} \\
 &=&G^{(0)\alpha\beta}(\B.r_1,\B.r'_1,\B.r_2)\delta (t-t_2) \ ,
 \nonumber
 \end{eqnarray}
 The series for $\B.\Sigma$ and $\B.\Phi$ can be resummed exactly, as
 is shown in Fig.~\ref{F5}. There is a price to pay: there are three
 new objects that appear as a result of this resummation, known as the
 `triple" vertices A, B and C. They differ in the number of wavy tails
 that connect them to the propagators. Note that there is no triple
 vertex with three wavy tails; such a vertex vanishes due to
 causality, as is discussed in Appendix A. Each of these vertices can
 be represented as an infinite series in terms of the same objects,
 i.e. the vertices A,B,C, and the correlator and propagator. The first
 diagrams in the infinite series for the triple vertices are shown in
 Fig.~\ref{F6}b.

In fact one is not limited to an infinite series representation. One
can also represent the triple vertices through exact and complete
resummations of the infinite series, but this is at the expense of
introducing yet another set of objects, this time of quartic nature.
There are four different quartic vertices, with one, two, three or
four straight ``legs". We denote them as $D_{m.n}$ with $m$ standing
for the number of wavy tails, and $n$ for the number of straight
tails, $m+n=4$. It is natural to discuss the various contributions
appearing in these objects as ``skeleton" and ``irreducible" and this
discussion is presented in Appendix A.

\begin{figure}
\epsfxsize=8.6truecm
\epsfbox{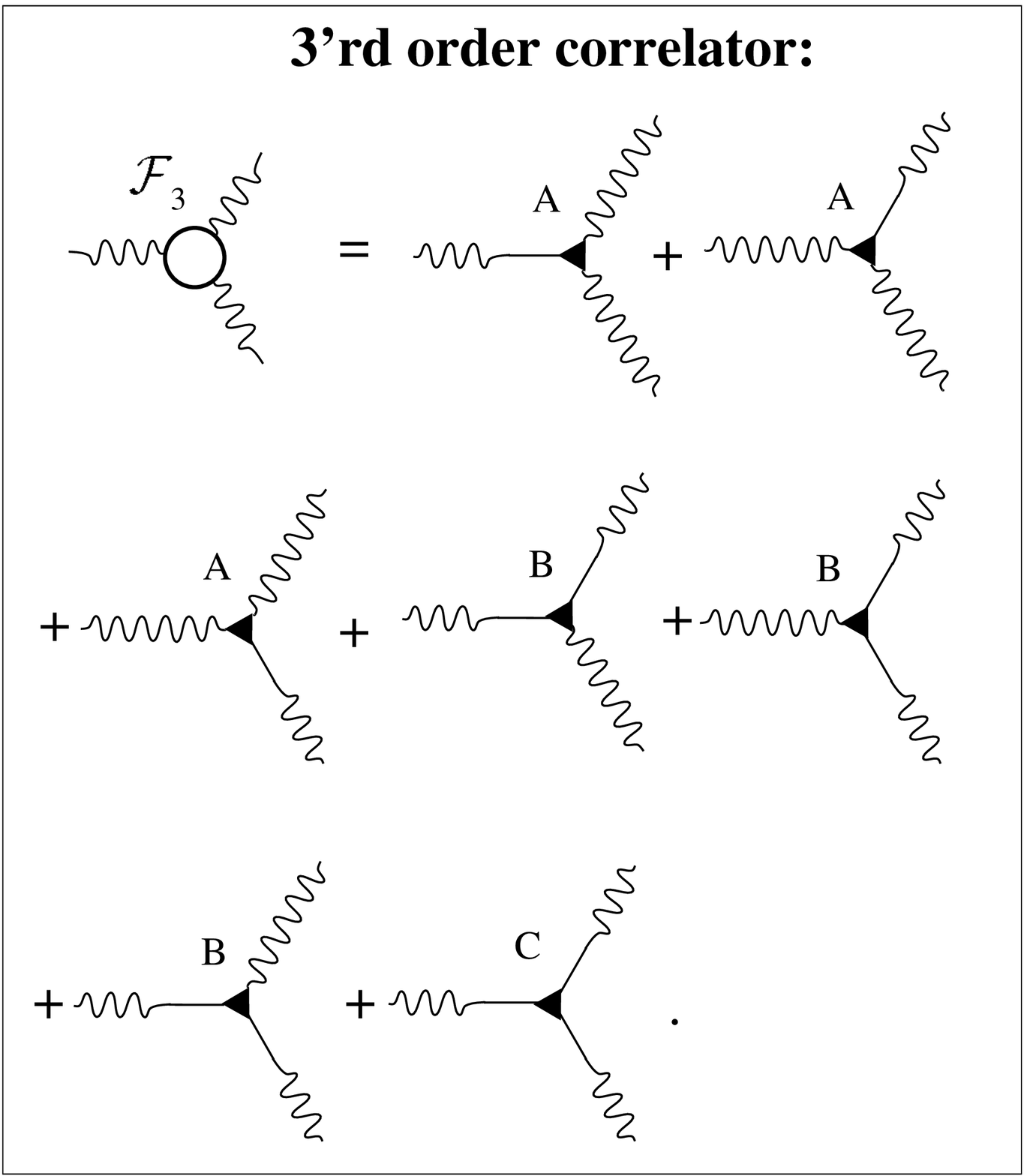}
\vskip 0.3cm 
\caption
{Exact representation of the third-order correlation function.}
\label{F8}
\end{figure}

\begin{figure}
  \epsfxsize=8.6truecm \epsfbox{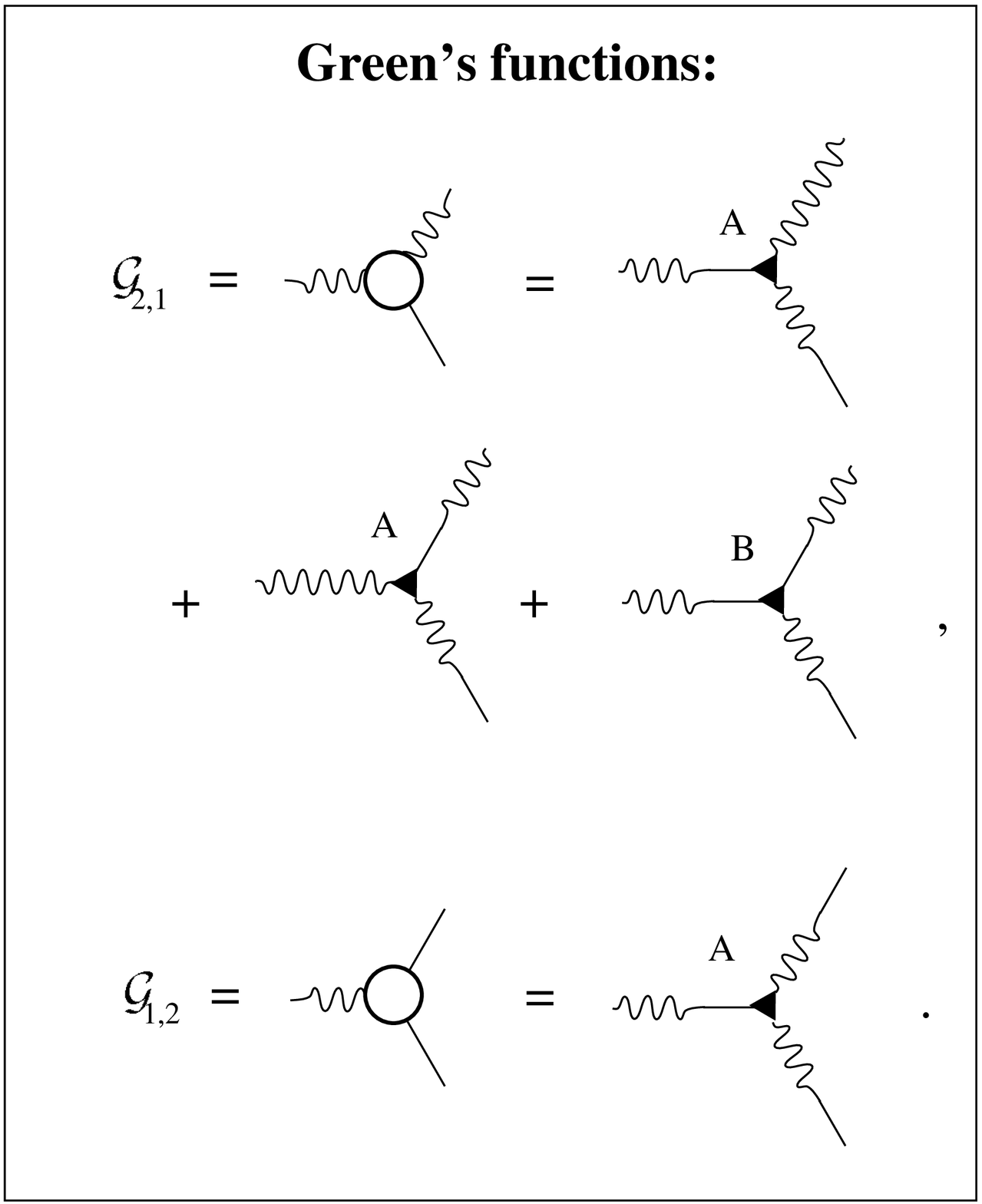} 
\vskip 0.3cm
\caption
{Exact representation of the nonlinear Green's functions ${\cal
    G}_{2,1}$ and ${\cal G}_{1,2}.$}
\label{F9}
\end{figure}

This process goes on. We can offer an exact, fully resummed equation
for the quartic vertices, but at the
price of introducing quintic objects,etc, as is explained in detail in
Appendix A. The renormalized vertices serve also in providing exact
representations for the many-point different-time correlation
functions (\ref{defFt}).  As examples we present in
Figs.\ref{F8},\ref{F9} the exact and fully renormalized three-point
correlation function $\B.{\cal F}_3$ and the nonlinear Green's
function $\B.{\cal G}_{2,1}$ in terms of the double correlator, the
Green's function and the triple vertices.

~~
\vskip-.8cm
 \begin{figure}
\epsfxsize=8.6truecm
\epsfbox{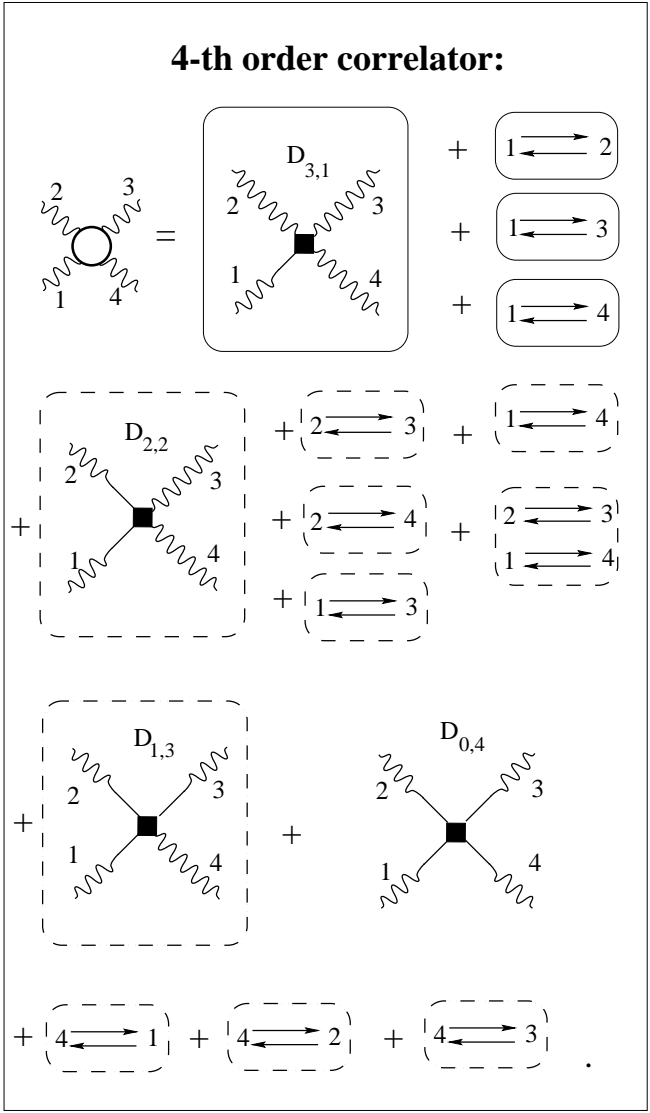}
\vskip 0.5cm 
\caption
{Exact representation of the 4th-order correlation function ${\cal
    F}_4$ in terms of 4th order vertices (black squares) and
  propagators.  The meaning of the 4th order vertex is discussed in
  Appendix A. The notation in this figure is that every contributions
  that is shown explicitly in a frame adds up to the indicated
  permutations enclosed in frames of the same line type.}
\label{F10}
\end{figure}

Similarly we can provide an exact representation of the four point
correlation function $\B.{\cal F}_4$ in terms of the same two-point
correlator and Green's function and the quartic vertices shown in Fig.
\ref{F10}. And so on; $\B.F_n$ can be represented as $2^n-1$
contributions in terms of n-legged vertices.

We see that we have a theory that allows exact representations of as
many renormalized objects as we desire. It appears clear and elegant,
but in fact is almost contentless. In order to endow it with
predictive power we need to learn how to perform two tasks. Firstly,
the space-time integrals are implicit in every diagram, see for
example Eqs.~(\ref{Dys},\ref{Wyl}). We need to learn how to compute
them. Secondly, we cannot deal with an infinite hierarchy, and we need
to learn how to close the hierarchy efficiently. In the next section
we review briefly the past history of attempts to solve these
equations, and why they failed in finding anomalous scaling.  This
will serve as a guidance for the new ideas that pave the 
route that is taken in the rest of this paper.
\section{Why is the Problem Still Open?}
The first serious attempt to derive the scaling exponents of the
statistical objects was due to Kraichnan \cite{59Kra,65Kra}, who
introduced the ``Direct-Interaction Approximation" (DIA).  In the
context of our presentation this approximation means truncating the
set of exact equations by replacing the dressed vertex A by its bare
counterpart, and by equating the dressed vertices B and C to their
bare value which is zero.  This approximation leads to closed
equations in terms of the second order correlation functions and
Green's function.  Kraichnan was the first to understand that after an
appropriate transformation of the velocity field (he chose Lagrangian,
but Belinicher and L'vov showed that their transformation amounted to
the same result) the space integrals in the DIA converge at both ends.
As a result of this ``locality" of the integrals, neither an inner nor
an outer scale appears in the bulk of the inertial interval. In
addition, Kraichnan assumed that the space-time correlation functions
are scale invariant in the sense of Eq.~(\ref{tscaleinv}).  This
assumption, together with the property of locality, lead to the
existence of scaling relations connecting the exponents $\zeta_2$ and
$z$:
\begin{eqnarray}\label{sc-rel1}
\zeta_2+2z&=&2\,,\\
2\zeta_2+z&=&2\ .\label{sc-rel2m
}
\end{eqnarray}
Of course, the solution is K41, i.e. $\zeta_2=z=2/3$. It should be
understood at this point that this disappointing result will continue
to hold not just at the DIA level, but at any finite order in the
dressing of the vertex via the perturbative scheme.  Since it was
shown that {\em every} diagram exhibits locality, the scaling
relations are reproduced by any finite truncation of the line resummed
perturbation theory.  Whenever we have a set of linear
scaling relations, K41 can be the only solution. It is always a
solution by dimensional reasoning, and therefore it must solve any set
of scaling relations. The linearity of the scaling relations is then
the end of the story.

One way out would be to assert that a full dressing of the vertex may
result in a new scaling exponent for that object. We still have a
linear set of two scaling relations, but now three exponents to
determine. The extra freedom allows deviations from K41.  However,
this is {\em not the case} as long as we believe that the scale
invariance as assumed in the form Eq.~(\ref{tscaleinv}) indeed holds.
It was shown by L'vov and Lebedev \cite{93LL} that under that
assumption the exponent of the vertex is unchanged from its bare value
of $-1$.  Accordingly, K41 seems a deeper trap than ever.  Is there a
way out?

Our contention is that the fundamental fallacy is the assumption
Eq.~(\ref{tscaleinv}).  One could think that there is a fundamental
problem with renormalized perturbation theory since the problem lacks
a small parameter.  To show that this is not the case we develop the
theory further, and show that the fully renormalized equations
obtained from resumming the diagrammatic expansions are exact and
equivalent to hierarchical equations that stem from the Navier-Stokes
equations. We will then argue that {\em finite} resummations of the
perturbation theory up to a given order will give successive
approximations which are useful for the solution of the fundamental
equations (\ref{greater}).

\section
{Equivalence of the fully renormalized diagrammatics and the exact 
hierarchical equations}

The student of fluid mechanics who is not familiar with field
theoretic methods may find the diagrammatic expansion discussed in
section 5 somewhat foreboding. The student of field theories, on the
other hand, may find them suspicious since the problem does not have a
small parameter. It is important therefore for general
understanding, and also crucial for the further development of our
theory, to show that the fully resummed equations discussed above are
equivalent to hierarchies of equations that can be derived exactly
from the Navier-Stokes equations.

Consider first the Dyson equations (\ref{Dys}). Applying to this
equation the operator $\Big[{\partial \over \partial
  t}-\nu(\nabla^2+\nabla_0^2)\Big]$ from the left, and using
Eq.~(\ref{b18}) we can write, in the limit $\nu \to 0$,
\begin{eqnarray}
 &&{\partial \over \partial t_1} {\cal G}^{\alpha\beta}_{1,1} 
 ( {\B.r}_{0}|X_1;x_2) = G^{(0)\alpha\beta}
 ({\B.r}_1, {\B.r}'_1,\B.r_2) \delta (t_1-t_2)  
 \label{b8} \\
 &&+  \int d {\B.r}_b
 \int_ {t_2}^{t_1} dt_b
 \tilde\Sigma^{\alpha\gamma}( {\B.r}_{0}|X_1;x_b)
 {\cal G}_{1,1} ^{\gamma\beta}({\B.r}_{0}|X_b;x_2) \ , \nonumber
  \end{eqnarray}
where
\begin{eqnarray}
&& \tilde\Sigma^{\alpha\gamma}( {\B.r}_0| X_1;x_b)=
\int d {\B.r}_a G^{(0)\alpha\delta} (\B.r_1,\B.r'_1,\B.r_a)
\label{sigtild} \\&& \times \Sigma^{\delta\gamma}( {\B.r}_{0}|
 {\B.r}_a, {\B.r}_a ,t_{1};x_b) \ . \nonumber
\end{eqnarray}

\begin{figure}
\epsfxsize=8.6truecm
\epsfbox{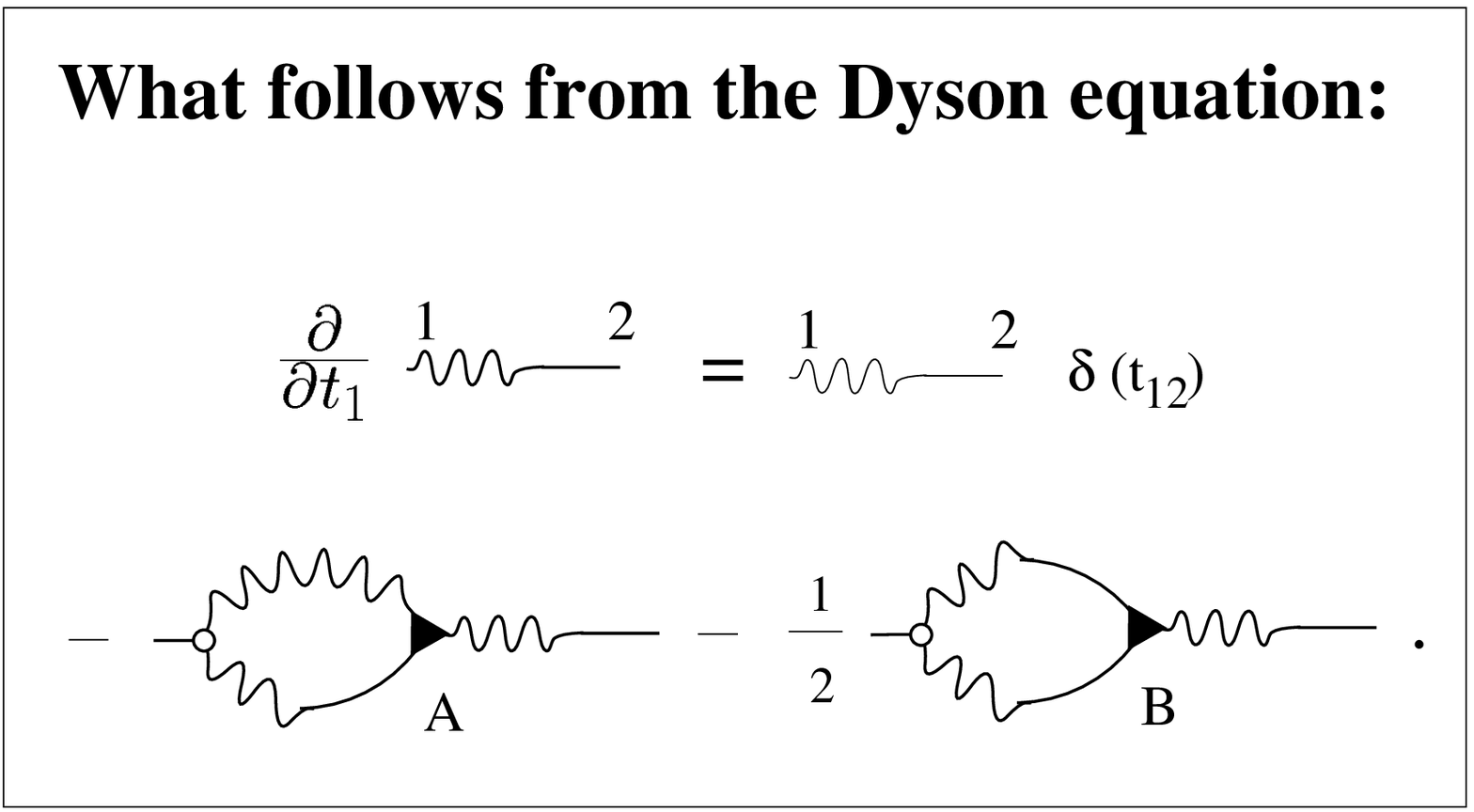}
\vskip 0.5cm 
\caption{The Dyson equation written in terms of
 the fully resummed vertices
A and B.}
\label{F11}
\end{figure}

Next one might compare Fig.~\ref{F11} with the exact representation of
the Green's function $\B.{\cal G}_{2,1}$ which is presented in
Fig.~\ref{F9}. The last term in Fig.~\ref{F11} is obviously a vertex
integrated against the last contribution to $\B.{\cal G}_{2,1}$ with
the B vertex, with a factor of $1/2$. The second term is just the
vertex integrated against the other two contributions with vertex A,
and in total we retrieve exactly the first equation in the hierarchy
presented in Fig.~\ref{F3}.

\begin{figure}
\epsfxsize=8.6truecm
\epsfbox{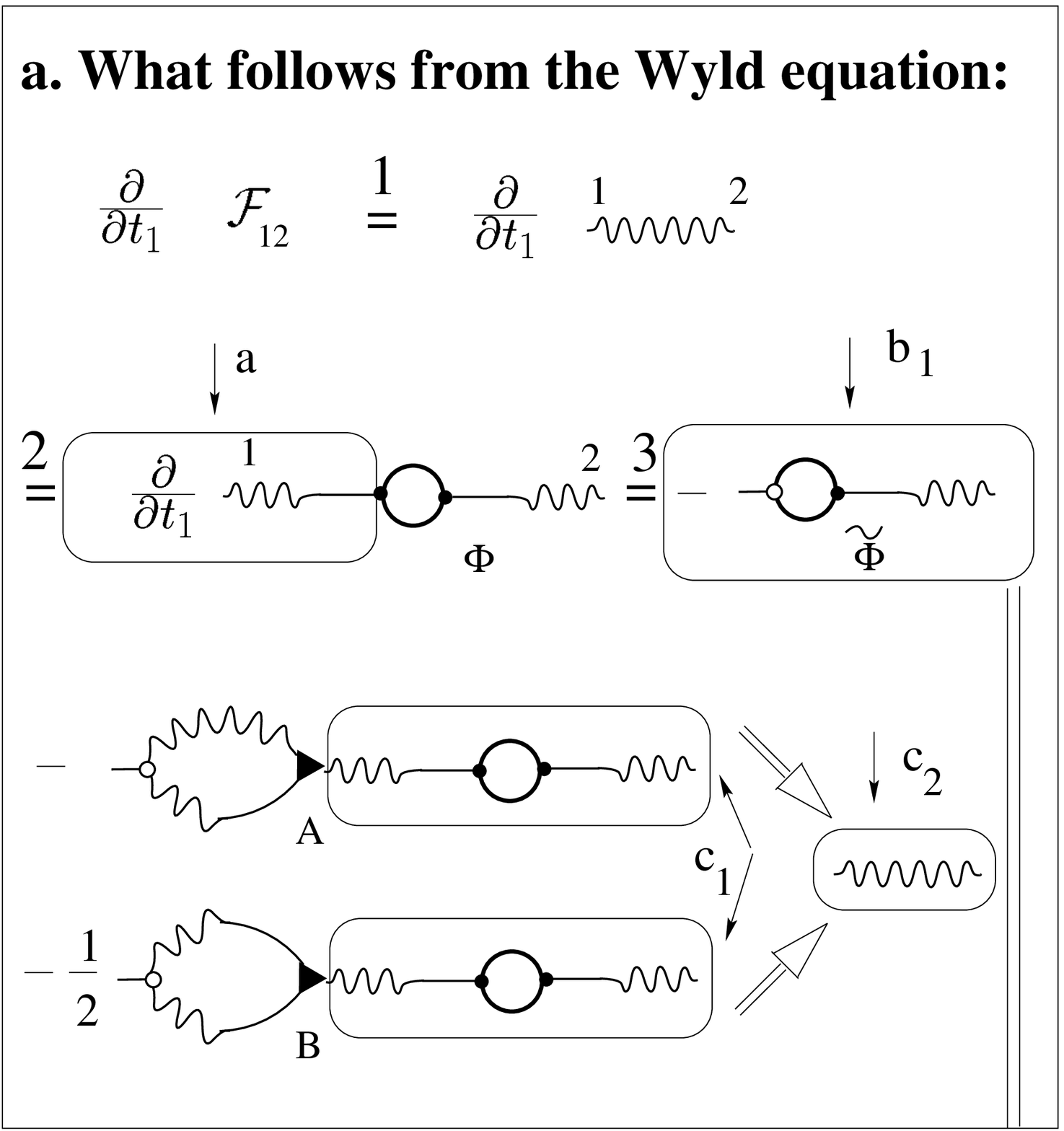}
\epsfxsize=8.6truecm
\epsfbox{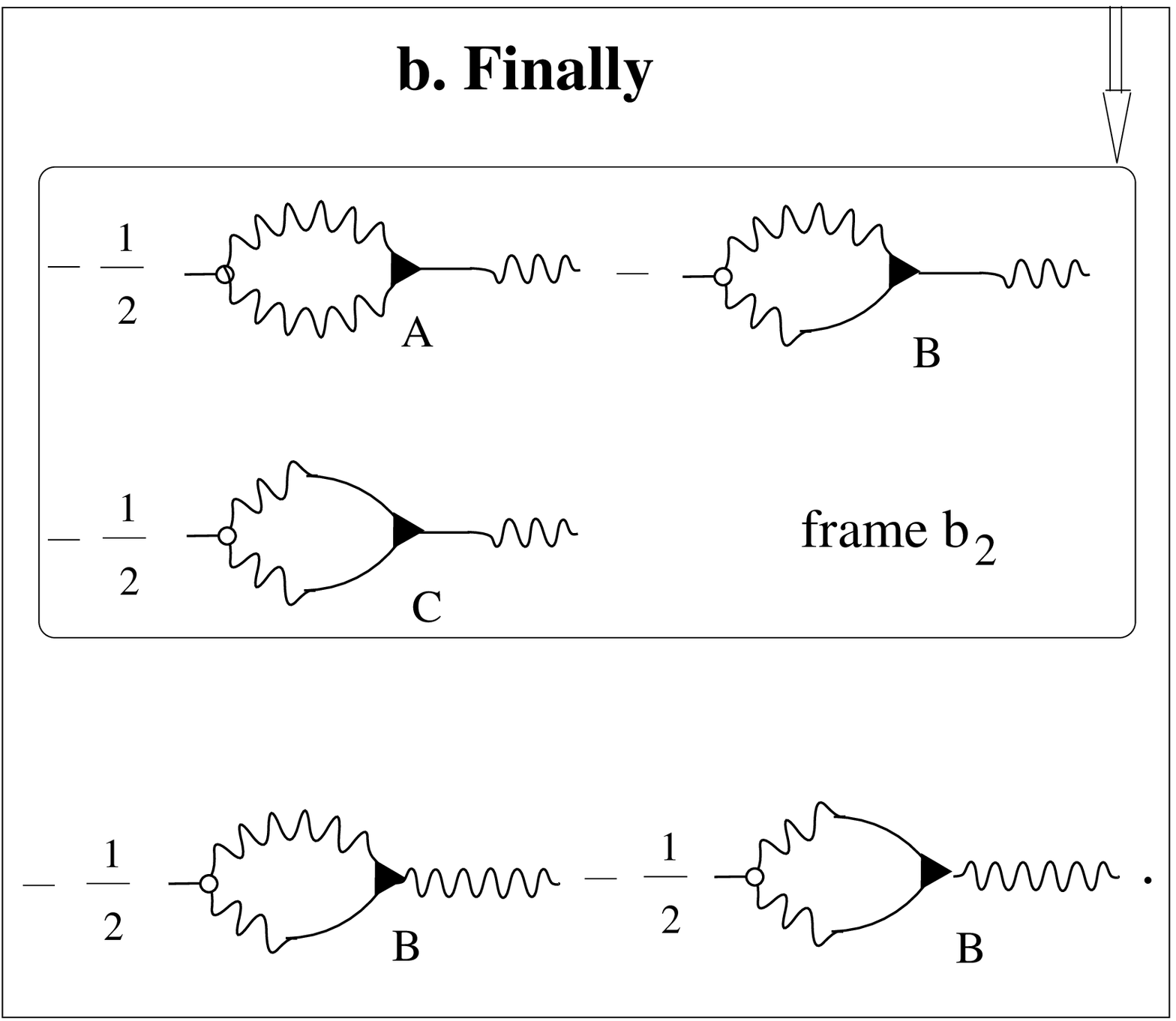}
\vskip 0.5cm 
\caption{Flow chart describing the diagrammatic derivation of the 
  first equation in the hierarchy of equations for the correlation
  functions.}
\label{F12}
\end{figure}

Figure \ref{F12} is a flow chart of the derivation of the first of the
hierarchy of equations for the correlation functions. The first
equation, denoted by 1 over the equal sign, is obvious.  The second
one, denoted by 2, is obtained using the Wyld equation (which is the
second of Fig.~\ref{F4}). Note that the time derivative is applied
only to the Green's function on the left. Equality 3 is a substitution
of the derivative of the Green's function from Fig.~\ref{F11}.  The
bare Green's function (the first term in Fig.~\ref{F11} changes the
dark circle to minus an empty one, in accordance with Eq(\ref{gG})).
One recognizes next that the pieces in the two frames designated $c_1$
are the 2nd order correlation function, shown in the frame $c_2$. The
first term shown in frame $b_1$ becomes the three contributions shown
in frame $b_2$ in panel b by using the exact representation of $
\B.\Phi$ shown in Fig.~\ref{F5} and relationship between vertices
$\Gamma$ and $\gamma$. Lastly, the equation in panel b contains all
the contributions to the RHS, which are nothing but the vertex
integrated against the 3rd order correlation function, as can be
checked by comparing with Fig.~\ref{F8}. We have thus derived the
first of the exact hierarchy of equations for the correlation
functions shown as the first line of Fig.~\ref{F2}.  One can proceed
similarly to derive the second of the hierarchic equations for the
correlation functions. The starting point would be the exact
representation of $\B.{\cal F}_3$ in Fig.~\ref{F8}. Taking the time
derivative with respect to $t_1$ one gets seven contributions, four of
which involve derivative of the Green's function $\B.{\cal G}_{1,1}$,
and three a derivative of $\B.{\cal F}_2$. For every such derivative
we need to substitute either the three terms appearing in
Fig.~\ref{F11}, or the five terms appearing in Fig.~\ref{F12}. Special
attention should be paid to the contribution of the first term (the
bare Green's function) in Fig.~\ref{F11}. When it is attached to one
of the dressed vertices A, B or C one needs to use their exact
representation as shown in Figs.\ref{F6}c. In all the other
contributions we can leave the dressed vertices A, B and C unchanged.
At this point we can collect all the terms, and find 60 terms
(counting each one with a factor 1/2 once and with a factor 1 twice).
One may then substitute the exact representation for the dark 4th
order vertices as a sum of four contributions, like those shown in
Fig.~\ref{F13}b in Appendix A. This will reduce the number of
contributions to 15, all with dark 4th order vertices.  This is
precisely the right number of contributions needed to reproduce the
exact result which is the empty circle vertex $\B.\gamma$ integrated
against the 15 contributions in $\B.{\cal F}_4$ shown in
Fig.~\ref{F10}. This completes the derivation of the second equation
in the hierarchy shown in Fig.~\ref{F2}. Exactly the same procedure
(but with fewer terms) provides the derivation of the second equation
in the hierarchy of the Green's function $\B.{\cal G}_{n,1}$ shown in
Fig.~\ref{F3} etc.

This derivation can be repeated at any order, showing the exact
correspondence of the fully resummed diagrammatics and the hierarchic
equations. Since the proof calls for the introduction of additional
graphical notations we present it in the appendix for the
consideration of the dedicated reader.

The summary of this section is as follows: The classical Dyson-Wyld
equations are equivalent to the first equations in the hierarchy of
equations for the correlation and Green's functions.  At this step of
resummation the 3rd order objects are given only in terms of an
infinite series in the 2nd order objects. If we replace at this point
the triple vertex by its bare value we will generate the Direct
Interaction Approximation. In the next step of resummation we have
2nd and 3rd order objects in fully resummed form. In particular the
3rd order vertices are not given in terms of infinite series but they
are exactly represented in a fully resummed form in terms of 4th
order objects as shown in Fig.~\ref{F6}c. At
this stage we can show agreement with the first two equations of the
hierarchies. The 4th order objects are still presented in terms of an
infinite series in terms 2nd and 3rd order objects, and one can
discuss various ways of approximating the 4th order objects. But one
can make instead the next step of resummation, that will yield fully
resummed 4th order objects in terms of 5th order ones, see
Fig.~\ref{F13}. The 5th order objects are represented at this stage as
an infinite series in terms of lower order objects. At this step we
can recover the first three equations of the hierarchies. This
procedure can be continued to any desired order, and at every $n$-th
step of the procedure in which we have fully resummed $n$-order
objects, we can recover the first $n-1$ equations of the hierarchies.
It is reasonable to assume that by deferring the closure
approximation to higher and higher order steps of the resummation we
may find better and better answers for the lower order objects. In the
next Section we will discuss several possibilities of making
intelligent closures.

\section{closure schemes}
In this section we explain why partly resummed perturbation expansions
are useful in implementing intelligent closures of the infinite
hierarchy (\ref{greater}). We take as an example a closure at the
level of the 3-point vertices A, B and C, but the comments are
relevant for closures at higher order as well. The complete discussion
belongs to the forthcoming paper \cite{97BLP}, and here we just
briefly sketch the important ideas, to underline why we spent so much
effort on the development of the resummed perturbation theory in
Sects. 5-7.

The main point is that we want to select the closure approximation
such as not to reintroduce power counting into the hierarchy
(\ref{greater}). Consider then the three point vertices A, B and C. We
have for each of these vertices an infinite series in terms of
themselves and the lower order 2-point correlation functions and
Green's functions, see Figs.\ref{F6}b.  Of course, taking all the
infinite contributions is impossible, and we need a criterion to
choose partial series of contributions as the approximation for the
3rd-order vertices. One traditional approximation that one could use
is the so-called ``triangular approximation" \cite{Tri}; this means
that we take all the triangular diagrams like the first four
contributions after the bare vertex in Fig.~\ref{F6}b. Each of these
triangular diagrams contains three dressed 3-point vertices of type A,
B or C, and three propagators (2-point correlator or Green's
function). These diagrams result from the infinite (partial)
resummation of all the ``planar" diagrams which do not contain crossed
lines of propagators. Note that the four shown triangular diagrams are
not all the diagrams of this type, but their number is small (for
example six in the case of vertex A).

In fact such an approximation is not suitable for our needs. In the
forthcoming paper \cite{97BLP} we will show that this approximation
reintroduces power counting to the hierarchy (\ref{greater}). On the
other hand we will show that the approximation that is obtained by
taking {\em only the skeleton contributions} and {\sl neglecting the
  fully irreducible contributions} is appropriate, since it does not
reintroduce power counting. In other words we can use exact
representations for the $n$-order fully irreducible vertices in terms
of partly reducible $n+1$-order vertex (as
shown in Fig.~\ref{F6} for 3'rd-order vertex and in Fig.~\ref{F13} for
the 4th order vertex) and then to neglect the fully reducible
contribution (empty objects) for $n+1$-order vertex in order to select that
appropriate infinite partial resummation that does not clash with the
rescaling symmetry (\ref{ressym}) that we want to preserve.  This will
be shown and utilized in the forthcoming paper \cite{97BLP}.  In
particular we will show that when we neglect the fully irreducible
contributions to the 4th order vertex we get closed nonlinear
equations for the triple vertices as shown in Figs.\ref{F6}c and
\ref{F7}.  By neglecting the fully   irreducible contributions to the
5th order vertex (see Figs.~\ref{F14}b as and example) we find closed
equations for the full 3rd and 4th order vertices, and so on.

\section{Appendix A: Discussion of the 4th and the 5th 
  order vertices} \vskip -.5cm There are four types of 4th order
vertices, as shown in Fig.~\ref{F7}a.  The various contributions to
the 4th order vertices are grouped according to their topology in
three classes as shown in Fig.~\ref{F7}b. The first class is known as
``two-eddy reducible" or ``skeleton" contributions. Such
contributions, denoted by joint triangles, can be split into two
pieces by cutting off one propagator, see panels c and d.  The second
class is known as ``irreducible", and is denoted as an empty square.
The grey square is the third class, termed "partly reducible", and
made of the irreducible vertex and the reducible contributions. The
need to consider separately the grey vetex stems from its appearance
in the exact representation of the third order vertices, see
Fig.\ref{F6}.  All the possible types of skeleton contributions to the
4th order vertices are shown in Figs.\ref{F7}. The diagrammatic
notation of the skeleton contribution is a mnemonic to stress the fact
that they consist of two dressed triple vertices which were denoted as
black triangles.  The joining at the apex hides a bridge that can be
made from any of the available propagators, either a correlator with a
wavy line, or a Green's function with a straight-wavy line with either
orientation. All these terms are presented in Fig.~\ref{F7}d. In
contrast, Fig.~\ref{F7}c have only one type of propagator, with the
straight line on the right.  The two other possibilities have wavy
lines on the right, and this requires a 3'rd order vertex with three
wavy tails which is zero identically. Note that the line resummed
theory does not have bridges made of two consecutive propagators with
something in between: such contributions are ``one-eddy reducible",
and all such diagrams have been already resummed in the Dyson line
resummation.

\begin{figure}
\epsfxsize=8.6truecm
\epsfbox{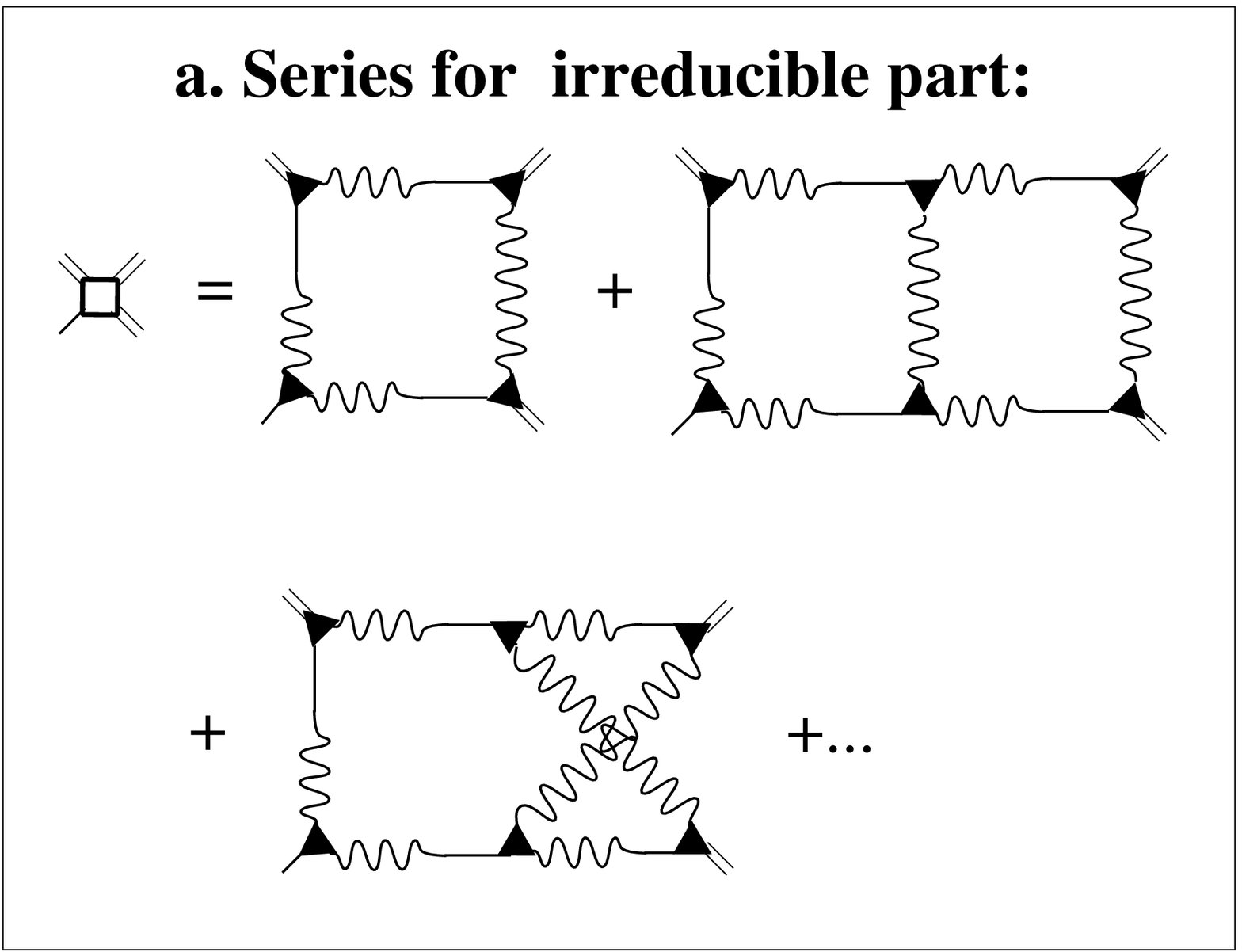}
\epsfxsize=8.6truecm
\epsfbox{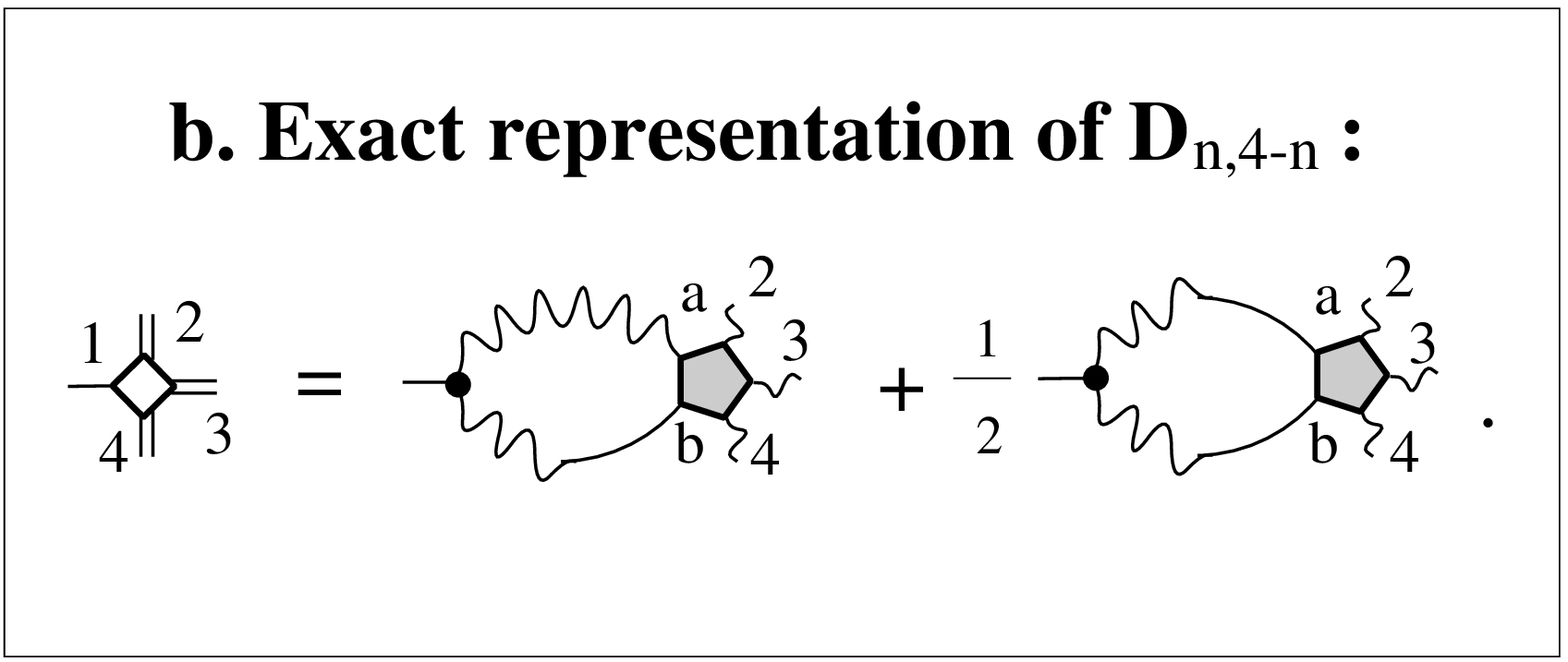}
 \vskip 0.5cm 
\caption
{Panel a: diagrammatic expansion of the quartic dressed vertices in
  terms of triple dressed vertices A, B, and C, and propagators.
  Panel b. Exact representation of the irreducible part of $D_{n,4-n}$
  via the partly reducible 5th order
  vertices shown in Fig~14b.  }
\label{F13}
\end{figure}~~

\begin{figure}
\epsfxsize=8.6truecm
\epsfbox{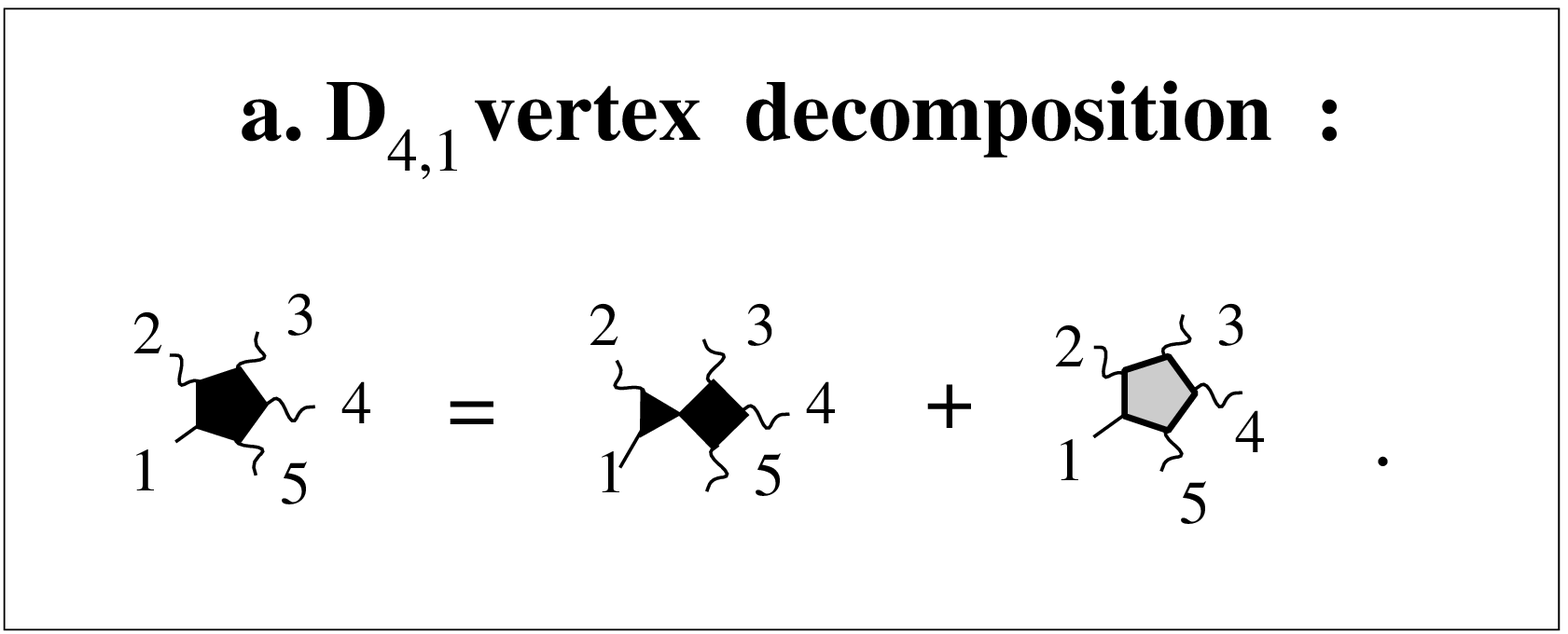}
\epsfxsize=8.6truecm
\epsfbox{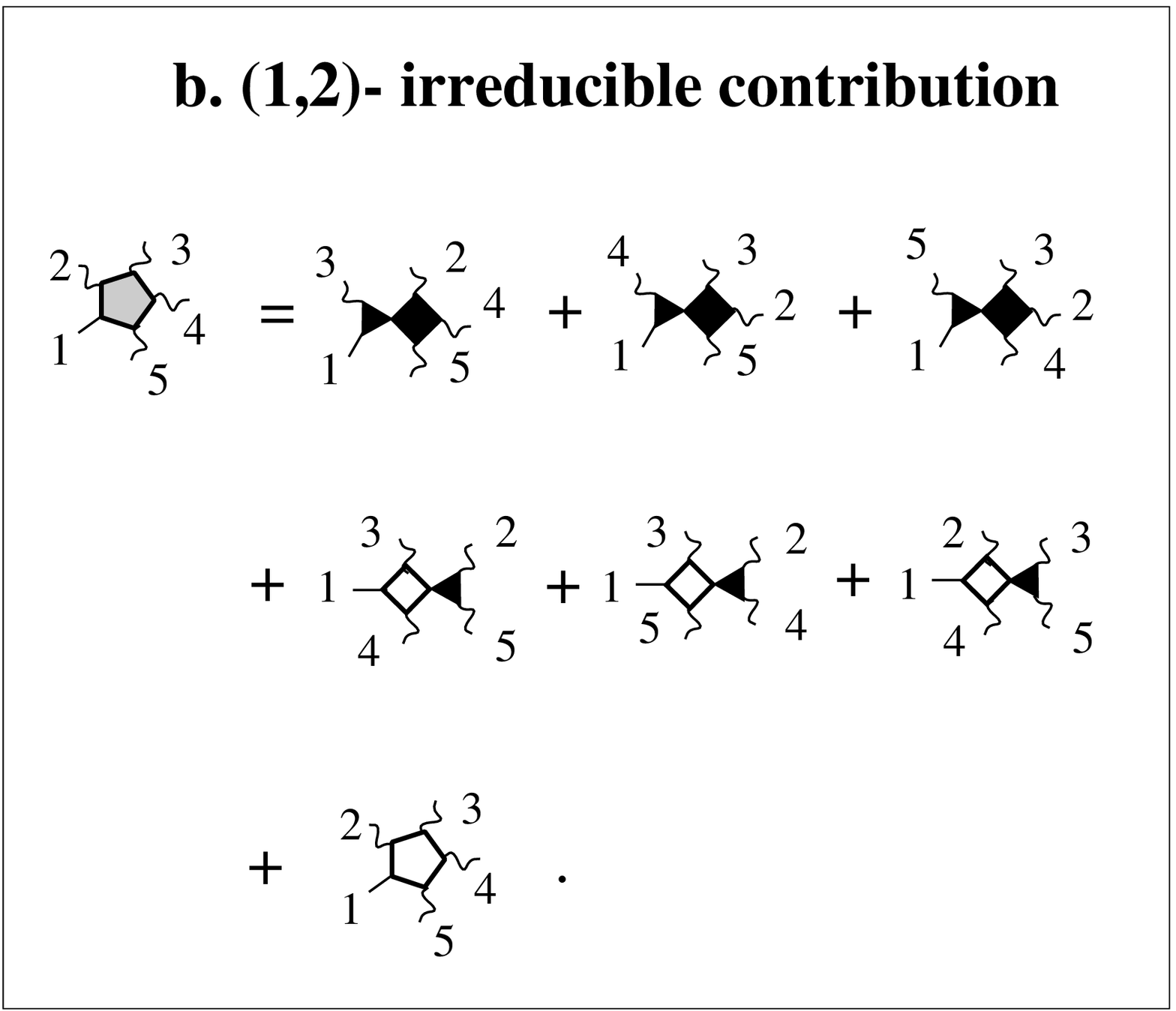}
\vskip 0.5cm 
\caption
{Decomposition of the 5th order vertex $D_{4,1}$.  Panel a:
  decomposition of $D_{4,1}$ into one skeleton contribution (first
  term) and partly reducible 5th order vertex (grey pentagon).  
Panel b: further decomposition of
  $D_{4,1}$ into reducible (skeleton) contributions (first three term)
and a fully irreducible contribution (empty pentagon). }

\label{F14}
\end{figure}

\begin{figure}
\epsfxsize=8.6truecm
\epsfbox{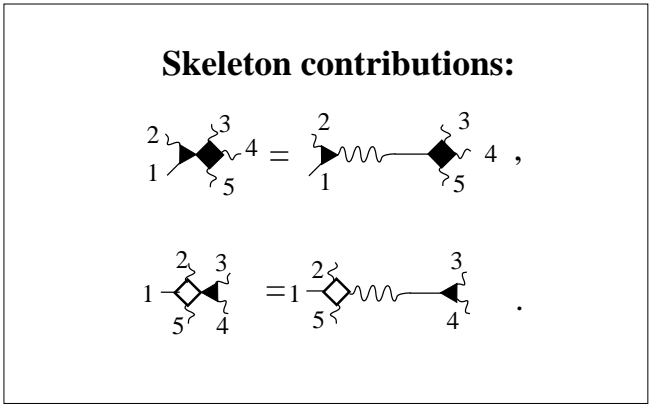}
\vskip 0.5cm   
\caption
{Examples of skeleton contributions to the 5th order vertex
$D_{4,1}$ .}
\label{F15}
\end{figure}

As noted, the skeleton diagrams are ``two-eddy reducible" in the sense
that they can be split into two pieces by cutting across the bridge.
Examples of irreducible diagrams for $D_{n,4-n}$ which cannot be cut
this way are shown in Fig.~\ref{F13}a. There are infinitely many of
them. Again, this kind of series can be resummed exactly, as is shown
in Fig.~\ref{F13}b. The price is the appearance of an $(n+1)$'th order
vertex, 5th order in the present case.  This vertex is denoted as a
grey pentagon. We will call it ``partly reducible" vertex.  Consider
now 5th order vertices, for example $D_{4,1}$, see Fig.~\ref{F14}. As
discussed above we distinguish here the reducible contribution which
is explicitly shown in Fig~.\ref{F14}a and the partly reducible
contribution, shown as gray pentagon. The need to use a special
notation (grey pentagon) for the partly reducible vertex is again
motivated by its appearance in the exact representatin of the 4th
order vertex, Fig.\ref{F13}b. Its further decomposition into reducible
parts and a fully irreducible term (empty pentagon) is shown in
Fig~.\ref{F14}b.
 Examples of reducible (skeleton) contributions to
a 5th order vertex are shown in Fig.~\ref{F15}.

In a forthcoming paper in collaboration with V. Belinicher we will
show that the skeleton contributions to the nth order vertices have
the same rescaling symmetry as the full nth order vertex. This
observation means that we can achieve a consistent closure by
neglecting the irreducible contributions at any given step of the
vertex resummation. It is therefore important to observe that the
skeleton contributions to the n'th order vertex are always made of
lower order vertices. As shown above, the skeleton contributions to
the 4th order vertices consist of two triple vertices and those for
the 5th order vertices have contributions of two types. Those with
full triple vertices connected by two propagator bridges and one full
triple vertex with one irreducible 4th order vertex connected by one
propagator bridge. We do not present here the skeleton contributions
to 6'th order vertices. There are four types of them: 1) Four black
triangles on a line, connected via three propagator bridges, 2) Four
black triangles in a star configuration with one central triangle
connected via three bridges with three other triangles, 3) Two
irreducible 4th order vertices (empty squares) connected by a
propagator, and 4) a triple vertex (black triangle) connected via a
propagator to a 5th order irreducible vertex.

\acknowledgments Some of the confusing aspects of the analysis
described above, and in particular the developments leading to
Eq.~(\ref{greater}) became much clearer due to discussions with Victor
Belinicher to whom we offer thanks.  This work was supported in part
by the German Israeli Foundation, the US-Israel Bi-National Science
Foundation, the Minerva Center for Nonlinear Physics, and the Naftali
and Anna Backenroth-Bronicki Fund for Research in Chaos and
Complexity.

\end{document}